\documentclass[apj]{emulateapj}
\usepackage{amsmath}
\usepackage{amssymb}
\usepackage{epsfig}
\usepackage{apjfonts}

\newcommand{\cms}{\,{\rm cm$^{-2}$}\,}
\newcommand{\cmc}{\,{\rm cm$^{-3}$}\,}
\newcommand{\kms}{\,{\rm km\,s$^{-1}$}\,}

\newcommand{\etal}{{ et~al.~}}

\newcommand{\sbasunit}{\,{\rm photons\,cm$^{-2}$\,arcsec$^{-2}$}\,}
\newcommand{\ergs}{\,{\rm erg\,s$^{-1}$}\,}

\newcommand{\Ms}{M_\odot}
\newcommand{\Zs}{Z_\odot}

\shorttitle{Merger of NGC\,6868 and NGC\,6861}
\begin{document}


\title{The Mysterious Merger of NGC\,6868 and NGC\,6861 in the 
Telescopium Group 
}
\author{M. E. Machacek, E. O'Sullivan, S.W. Randall, C. Jones, and 
    W. R. Forman}
\affil{Harvard-Smithsonian Center for Astrophysics \\ 
       60 Garden Street, Cambridge, MA 02138 USA
\email{mmachacek@cfa.harvard.edu}}

\begin{abstract}
We use {\it Chandra} X-ray observations of the hot gas in and 
around NGC\,6868 and NGC\,6861 in the Telescopium galaxy group 
(AS0851) to probe the interaction history between these galaxies. 
Mean surface brightness profiles for NGC\,6868 and NGC\,6861 are each
well described by double $\beta$-models, suggesting that they are
each the dominant galaxy in a galaxy  subgroup about to merge.
Surface brightness and temperature maps of the 
brightest group galaxy NGC\,6868 show a cold front edge $\sim 23$\,kpc 
to the north, and a cool $0.62$\,keV spiral-shaped tail to the 
south. Analysis of the temperature and density  
across the cold front constrains the relative motion between 
NGC\,6868 and the ambient group gas to be at most transonic; while  
the spiral morphology of the tail strongly suggests that the cold
front edge and tail are the result of gas sloshing due to the subgroup
merger. The cooler central region of NGC\,6861 is surrounded by a sheath
of hot gas to the east and hot, bifurcated tails of X-ray emission to the west
and northwest. We discuss supersonic infall of the NGC\,6861 subroup, 
sloshing from the NGC\,6868 and NGC\,6861 subgroup merger, and AGN 
heating as possible explanations for these features, and 
discuss possible scenarios that may contribute to the order of 
magnitude discrepancy between the Margorrian and 
black hole mass - $\sigma$  predictions for its central black hole.  

\end{abstract}



\section{INTRODUCTION}
\label{sec:introduction}

One of the most important questions facing models of galaxy evolution 
today is how central supermassive black holes, found at the centers of 
most galaxies, co-evolve with their host galaxies.  The observed 
correlations between central black hole mass and stellar velocity 
dispersion in present epoch galaxies 
($M_{\rm BH} - \sigma_*$, Gebhardt \etal 2000; 
Ferrarese \& Merrit 2000; Tremaine \etal 2002) and between black hole 
mass and galaxy bulge mass or bulge luminosity 
($M_{\rm BH} - M_{\rm bulge}$; Magorrian \etal 1998;  H\"{a}ring \&
Rix 2004) provide striking evidence for this 
coevolution. However, the shapes of these correlations, particularly
at the extremes of velocity dispersion or luminosity, are not well
known, and, which, if either, of these correlations is a  reliable
predictor of black hole mass in galaxies with high central stellar
velocity dispersion or large luminosity remains controversial 
(Lauer \etal 2007).  Key to resolving this controversy is
understanding the dynamical connections between 
galaxy interactions, the feedback cycle from  
active galactic nuclei (AGNs), and black hole fueling and growth. 
 
Recent observations show that galaxy evolution occurs 
predominantly in moderately massive galaxy groups at 
redshifts $\gtrsim 1$ (Cooper \etal 2006). Thus to 
build dynamical models for the transformation of galaxies and the
growth of their central black holes, we must understand how galaxy 
interactions and feedback affect gas in and surrounding galaxies in
the group environment.  Questions that remain poorly understood include: 
What dynamical processes control the mass and energy flows into or out 
of the merging galaxies either fueling the central AGN and/or
initiating star formation,  or quenching star formation and heating
and enriching the intragroup medium (IGM)?  When and how are the 
dark matter and hot gas halos of the merging
galaxy separated from the galaxy to become part of
the group gravitational potential? How efficiently are the outer
stellar halos of interacting galaxies stripped? Are central stellar 
velocity dispersions of the galaxies  
affected by these interactions and on what dynamical timescales? What 
role do non-hydrostatic group gas (IGM) motions, induced by galaxies 
and galaxy subgroups passing through the group core, play in the 
co-evolution of the system?

Merging galaxies in nearby galaxy groups are particularly important 
laboratories in which to address these questions, since only in these 
groups do X-ray observations have sufficient angular resolution to 
directly observe the X-ray edges, outflows, cavities, shocks, ripples 
and tails characteristic of the dynamical processes at work on the 
galaxies and within the group. Measurements of temperatures and
densities in these features allow us to constrain the 
velocities, orbits and interaction history of the galaxies, as well as 
probe the cycle of nuclear activity that transports  
matter and energy between the galaxy's nucleus and the 
surrounding gas. Modeling the effects of these
interactions on nearby galaxies, that lie on the extremes of the 
black hole mass scaling relations, provides a unique opportunity to
understand the galactic processes that determine the shape and validity 
of these relations. 

\begin{figure}[t]
\begin{center}
\includegraphics[height=1.722in,width=3.0in]{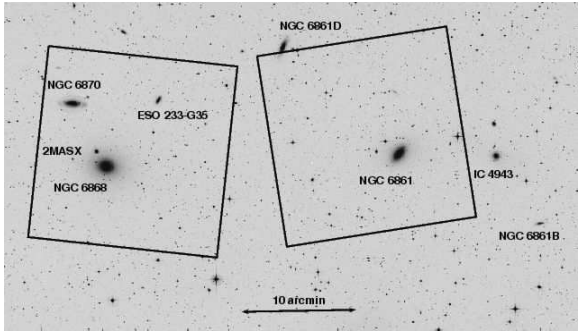}
\caption{\footnotesize{Bj-band image of the Telescopium (AS0851) group with the 
brightest group galaxy NGC\,6868 to the left and second brightest 
group galaxy NGC6861 to the right. Other bright members of AS0851 are 
also labeled. Squares show the {\it Chandra} ACIS-I fields-of-view 
(FOVs) for the observations discussed in this work. North is up and
East is to the left.}}
\label{fig:dss}
\end{center}
\end{figure}

In this paper we present the results of 
{\it Chandra} observations of the two brightest elliptical galaxies, 
NGC\,6868 and NGC\,6861, in the Telescopium group (AS0851). 
AS0851, shown in Figure \ref{fig:dss}, is an Abell 
class $0$ galaxy group $28$\,Mpc distant
(Tonry \etal 2001), containing $ > 11$ member galaxies (Garcia 1993). 
Both the brightest group member, elliptical galaxy NGC\,6868 
($20^h09^m54.1^s$,$-48^\circ22'46\farcs4$, $B_T=11.66$, NED) 
and the second brightest group galaxy, SA0 galaxy NGC\,6861
($20^h07^m19.5^s$,$-48^\circ22'12\farcs8$,$B_T = 12.12$, NED), 
$25\farcm7$ to the west, show signs of recent interactions 
and are likely to merge. Surprisingly, Wegner \etal (2003) measured 
a central stellar velocity dispersion of the less luminous galaxy (NGC\,6861)  
of $\sim 414$\kms, a factor $1.6$ greater than the $250$\kms central 
stellar velocity dispersion of the more luminous elliptical galaxy
NGC\,6868. 

We summarize the galaxy properties and estimates of the central black
hole masses for both NGC\,6868 and NGC\,6861 in Table \ref{tab:bhmass}.
Using the $250$\kms central stellar velocity 
dispersion for NGC\,6868 in the black hole mass - stellar velocity 
dispersion ($M_{BH}-\sigma_*$) relation, 
\begin{equation}
 {\rm log}\,\Big (\frac{M_{BH}}{\Ms}\Big ) = (8.13 \pm 0.06) + 
    (4.02 \pm 0.32)\, {\rm log}\Big ( 
      \frac{\sigma_*}{200\,{\rm km\,s}^{-1}}\Big ),
\label{eq:mbhsigma}
\end{equation} 
where $M_{BH}$ is the mass of the black hole and $\sigma_*$ is the
central stellar velocity dispersion (Tremaine \etal 2002; 
Ferrarese \& Merrit 2000; Gephardt \etal 2000), 
we infer a central black hole mass in NGC\,6868  of $\sim 3.3 \times
10^8\Ms$, typical for black hole masses in elliptical galaxies in 
groups (see, e.g. Tremaine \etal 2002, Verdoes Kleijn \etal 2006).  
An alternative estimator, the Magorrian relation, relates the mass 
of the central black hole $M_{BH}$ to the stellar mass $M_*$ of the 
galaxy's bulge (as stated in H\"{a}ring and Rix 2004; see also 
Magorrian \etal 1998; Marconi \& Hunt 2003), 
\begin{equation} 
 {\rm log}\,\Big (\frac{M_{BH}}{\Ms}\Big ) = (8.20 \pm 0.10) + (1.12 \pm
   0.06) {\rm log}\, \Big (\frac{ M_*}{10^{11}\Ms}\Big ). 
\label{eq:magorrian}
\end{equation}
For elliptical and spheroidal galaxies the bulge mass is replaced with 
the total stellar mass of the galaxy. Following Gilfanov (2004), we use the 
2MASS total K-band luminosity  of NGC\,6868 to estimate the total
stellar mass for NGC\,6868 in eq. \ref{eq:magorrian} 
and find $M_{BH} = 3.1 \times 10^8\Ms$, in excellent agreement with 
the $M_{BH}-\sigma_*$ relation. 

Optical measurements of NGC\,6868 reveal a central dust lane with weak 
spiral features (Buson \etal 1993, Veron-Cetty \& Veron 1988, Hansen
\etal 1991), and indicate that gas at the center of the galaxy 
is not moving in regular orbits (Zeilinger \etal 1996). 
IRAS measurements of NGC\,6868 at $60$ and $100\,\micron$ confirm 
the presence of cold dust in the galaxy (Bregman \etal 1998). 
A possible explanation for these dust features and 
unusual stellar kinematics is that NGC\,6868 has recently captured a 
dust rich companion. Such a cold gas rich merger could fuel AGN
activity and/or black hole growth (Hardcastle \etal 2007; Machacek \etal
2008).  NGC\,6868 has been observed in the radio at 
$2.3$, $5$, and $8.4$ GHz with the Parkes and Australian Telescopes 
(Slee \etal 1994) and at $843$ MHz in the SUMSS survey (Mauch \etal
2003). The galaxy is classified as a low-power, flat spectrum ($\alpha
\sim 0.07$) radio source  with total flux densities of $139$\,mJy,
and $124$\,mJy at $843$\,MHz and  $5$\,GHz, respectively. The brightness
temperature and spectral slope are inconsistent with thermal emission
from star forming HII regions, such that the source of the radio
emission is likely an active galactic nucleus.  

On the other hand, the  $\sim 400$\kms 
central velocity dispersion for
NGC\,6861, located $\sim 206$\,kpc 
away from NGC\,6868, is one of the highest measured for any 
early type galaxy (Wegner \etal 2003; 
Koprolin \& Zeilinger 2000). 
From the $M_{BH}-\sigma_*$ relation (eq. \ref{eq:mbhsigma}), 
we infer a supermassive black hole in
NGC\,6861 of $\sim 2.5 \times 10^9\Ms$, similar to that in M87, 
the central dominant giant elliptical galaxy in the Virgo galaxy 
cluster, and to that in NGC\,4649, a giant elliptical galaxy also 
in Virgo (Tremaine \etal 2002). This is an order of magnitude higher
than the $2.1 \times 10^8\Ms$ black hole mass  expected for NGC\,6861 
from the Magorrian relation (eq. \ref{eq:magorrian}).

The dispersion velocity for gas in the inner $30-50$\,pc of NGC\,6861, 
close to the black hole sphere of influence, has been measured using
central emission line widths from Hubble Space Telescope STIS data 
(Verdoes Kleijn \etal 2006; Beifiori \etal 2009). Both groups confined
their analysis to galaxies that did not show unusually disturbed
stellar kinematics. In each case the measured central gas velocity for 
NGC\,6861 is among the highest in their samples, comparable or
exceeding that found in NGC\,4486 (M87). However, extracting the 
black hole mass from the measured gas dispersions is highly sensitive
to the assumed morphology of the gas and, for the thin disk models used by
these authors, the unknown inclination of the central gas disk. 
Beifiori \etal (2009) argue that the data are consistent with 
a range of black hole masses, i.e. $\sim 1.5 \times 10^9\Ms$ for
disk inclination angle $i = 33^\circ$ to $3.6 \times 10^8\Ms$ for 
$i=81^\circ$, that within the uncertainties of the scaling relations, 
span the predictions of either the Magorrian or $M_{BH}-\sigma_*$ relation. 

These data suggest several possible scenarios. The black hole mass for
NGC\,6861 may be high, as suggested by the $M_{BH}-\sigma_*$ relation,
and NGC\,6861 may itself be the central galaxy in a massive subgroup 
dark matter halo merging with the NGC\,6868 subgroup. 
Alternatively, the high central gas velocity dispersion in NGC\,6861 
could be caused, in part, by the input of kinetic energy into the gas
from the central AGN through outflows or turbulence 
(Verdoes Kleijn \etal 2006; Capellari \etal 2000). Then the mass of
the central black hole needed to explain the high central gas
velocities could be smaller, falling below 
the $M_{BH}-\sigma_*$ prediction and closer to that of the Magorrian
relation (eq. \ref{eq:magorrian}). NGC\,6861 does host  a weak radio
source, with core radio power, measured with the Australian Telescope,
of $6$\,mJy at $5$\,GHz (Slee \etal 1994). 
The radio source is extended with a core-to-total radio power ratio 
of $\sim 0.25$. Although there is no evidence for large scale
radio jets, the $5"$ ($0.7$\,kpc) angular resolution of the Australian 
Telescope does not rule out substructure on scales of a few tens of parsecs
nor preclude feedback from previous episodes of AGN activity, each of 
which might heat the central interstellar medium (ISM), lowering the 
black hole mass required to model the central gas velocity dispersions
observed with STIS. Finally, the stellar velocity dispersion of
NGC\,6861 could be anomalously high due to gravitational interactions 
with NGC\,6868 or other group galaxies. In these latter
two cases, the puzzle is then what 
interaction history could cause such a high stellar velocity
dispersion for NGC\,6861, while not producing obviously unrelaxed 
stellar kinematics that would have excluded the galaxy from the 
Beifiori \etal (2009) and Verdoes Kleijn \etal (2006) samples. 

NGC\,6868 and NGC\,6861 have been observed in X-rays with both the {\it
  ROSAT} and {\it Chandra} X-ray Observatories. These 
observations have been used to measure total X-ray luminosities and
test X-ray scaling laws (Beuing \etal 1999, O'Sullivan \etal 2001, 
Ellis \& O'Sullivan 2006) and also to determine average density and 
temperature profiles for the galaxies (Fukazawa \etal 2006). 
In this paper we use the {\it Chandra} X-ray measurements of the density
and temperature distribution of hot gas in NGC\,6868, NGC\,6861, 
and in the AS0851 group core to investigate the interaction history of these
galaxies and how those interactions may affect nuclear 
activity and black hole growth. Our discussion is organized as follows:  
In  \S\ref{sec:obs} we review the {\it Chandra} observations and our 
data reduction and processing procedures.
In \S\ref{sec:meanprop} we  discuss the mean X-ray surface brightness 
around each galaxy and search for substructure in the AS0851 group.
In \S\ref{sec:interact} we analyse density and temperature features
observed in these data and how they may constrain possible interaction
scenarios between the galaxies and group IGM. 
In \S\ref{sec:discuss} we comment on how these interactions may impact
the black hole mass relations. We summarize our results 
in \S\ref{sec:conclude}. 
Unless otherwise indicated, quoted uncertainties are 
$90\%$ confidence levels for spectral
parameters and $1\sigma$ uncertainties for X-ray counts. 
Coordinates are J2000. We adopt 
a luminosity distance of $28$\,Mpc, obtained from surface
brightness fluctuation measurements of NGC\,6868 (Tonry \etal 2001), 
as representative of the distance to the core of the 
AS0851 galaxy group and to NGC\,6861. Assuming the five year WMAP values for 
cosmological parameters  
($\Omega_m=0.27$, $\Omega_{\lambda} = 0.73$, ${\rm H}_0 = 0.71$;
Hinshaw \etal 2009), $1''$ at $28$\,Mpc corresponds to a distance 
of $0.134$\,kpc in the plane of the sky. 


\section{{\it CHANDRA} OBSERVATIONS AND DATA REDUCTION}
\label{sec:obs}

NGC\,6868 was observed for $23.5$\,ks on 2002 January 11 (Obsid 3191) 
and NGC\,6861 for $22.8$\,ks on 2002 July 26
(Obsid 3190), each with the Advanced CCD Imaging Spectrometer -
Imaging Array (ACIS-I; Gamire \etal 1992, Bautz \etal 1998) in VFAINT
mode on board the {\it Chandra} X-ray Observatory. We used the 
standard X-ray analysis packages from CIAO3.4 and FTOOLS for image
processing and spectral extraction and XSPEC11.3.0 for spectral 
modeling. The data were initially filtered to reject bad grades
($1$,$5$,$7$) and data that fell on hot pixels. Data flagged in
VFAINT mode as having excessive count rates in the border pixels
surrounding event islands were also removed to optimize signal to
noise at energies below $1$\,keV, where most of the source emission is
expected for these galaxies. We then reprocessed the data and created 
exposure maps and response files applying the most recent gain tables 
and instrumental corrections, including corrections for the charge
transfer inefficiency on the ACIS-I CCDs, the time-dependent build-up
of contaminants on the optical filter, and the secular drift of the
average pulse-height amplitude for photons of fixed
energy.\footnote{see, http://cxc.harvard.edu/contrib/alexey/tgain/tgain.html}   
The data were cleaned using a $3\sigma$ clipping algorithm in the
script lc\_clean to remove periods of anomalously high and low count rates.
Inspection of the resulting light curves showed that this was 
adequate for OBSID 3191, resulting in a useful exposure of $21551$\,s
for that observation. However, inspection of the light curves for 
OBSID 3190 (NGC\,6861) showed that 
contamination from a strong particle flare (ringing) persisted
throughout the latter half of the observation. Thus we used only the 
pre-flare data from OBSID 3190 in our analysis, resulting in a useful
exposure of $9852$\,s for NGC\,6861. 

Backgrounds for the imaging analysis and for spectral analysis of the
IGM were constructed from the $1.5 \times 10^6$\,s source free 
background data set\footnote{see
  http://cxc.harvard.edu/contrib/maxim/acisbg} appropriate for our
observation dates and instrument configuration. Background data
sets for each observation were normalized by comparing count rates in 
the $9.0-11.5$\,keV energy band, where particle background dominates. 
This resulted in additional normalization factors of $0.98$ and $1.07$ 
for the backgrounds for OBSID 3191 and OBSID 3190, respectively.
 We identified X-ray point sources in each observation 
in the $0.3-7$\,keV energy band using a wavelet decomposition
algorithm with a $5\sigma$ detection threshold. Using $10$ photon
counts, we identify point sources to limiting $0.3-7$\,keV fluxes of $8.6
\times 10^{-15}$\ergs and $4.3 \times 10^{-15}$\ergs for OSID 3190
(NGC\,6861) and OSID 3191 (NGC\,6868), respectively.
Since we are interested in the properties of the diffuse X-ray emission and
possible AGN activity, X-ray point sources, other than the galaxies' 
nuclei, were removed from all subsequent analyses. 

\begin{figure}[t]
\begin{center}
\includegraphics[height=1.404in,width=3in]{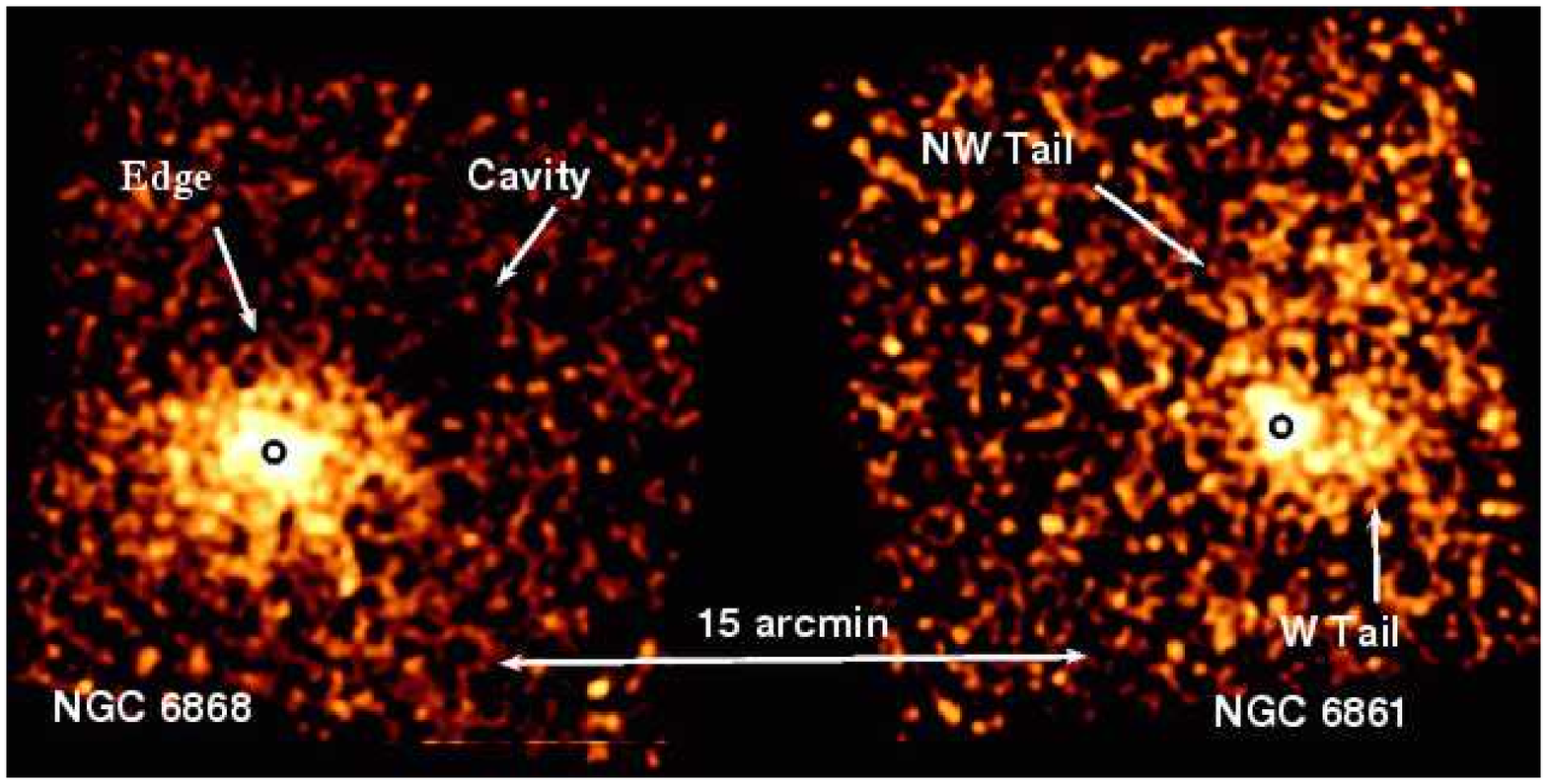}
\includegraphics[height=1.452in,width=3in]{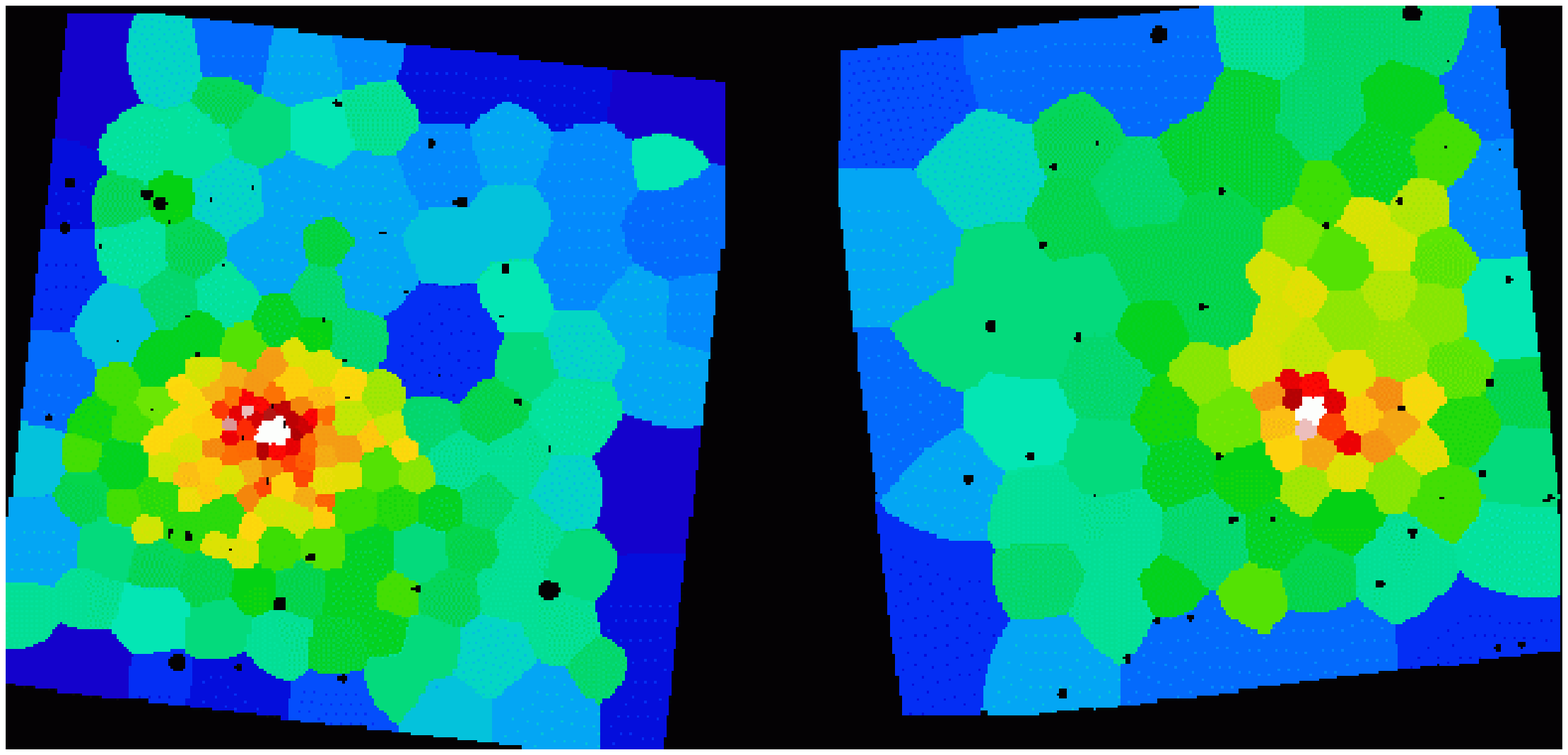}
\caption{\footnotesize{(upper) $0.5-2$\,keV {\it Chandra} mosaic of 
diffuse emission in
 NGC\,6868 and NGC\,6861 in the AS0851 galaxy group. Point sources
 have been excluded and the resulting image has been
  background-subtracted, exposure corrected and smoothed with an $8"$
  Gaussian kernel. $1 {\rm pixel} = 1\farcs968 \times 1\farcs968$. 
  Circles indicate the galaxy centers. North is up and east is to the
  left.
  ({\it lower)} $0.5-2$\,keV image of NGC\,6868 and
  NGC\, 6861 adaptively binned using Voronoi tessellation, with
  $3\farcs936 \times 3\farcs936$ pixels and  $S/N = 5$, to highlight
  the fainter structures. 
}}
\label{fig:chandramosaic}
\end{center}
\end{figure}

\section{Mean Surface Brightness Profiles and Group Structure}
\label{sec:meanprop}

In the upper panel of Figure \ref{fig:chandramosaic} we show a 
mosaiced  $0.5-2$\,keV Chandra image of the diffuse emission 
surrounding NGC\,6868 (left) 
and NGC\,6861 (right) in the galaxy group AS0851. The image has been
background subtracted and exposure corrected, and, after exclusion of
X-ray point sources, has been smoothed with an $8"$ Gaussian kernel. 
In the lower panel of Figure \ref{fig:chandramosaic} we present an
adaptively binned $0.5-2$\,keV Chandra image of the same regions,
using the Voronoi tessellation technique with a signal-to-noise ($S/N$) of
$5$ to highlight faint emission features.  
We first construct mean surface brightness profiles for 
NGC\,6868, NGC\,6861 and the surrounding gas to understand the
structure of the group as a whole and search for evidence of
substructure in AS0851 that might signal the merger of two subgroups. 
In the following section (\S\ref{sec:interact}), we will  
discuss the evidence for interactions suggested by the 
surface brightness features identified in Figure \ref{fig:chandramosaic} 
in and around each galaxy.
\begin{figure}[t]
\begin{center}  
\includegraphics[height=3.0in,width=2.37in,angle=270]{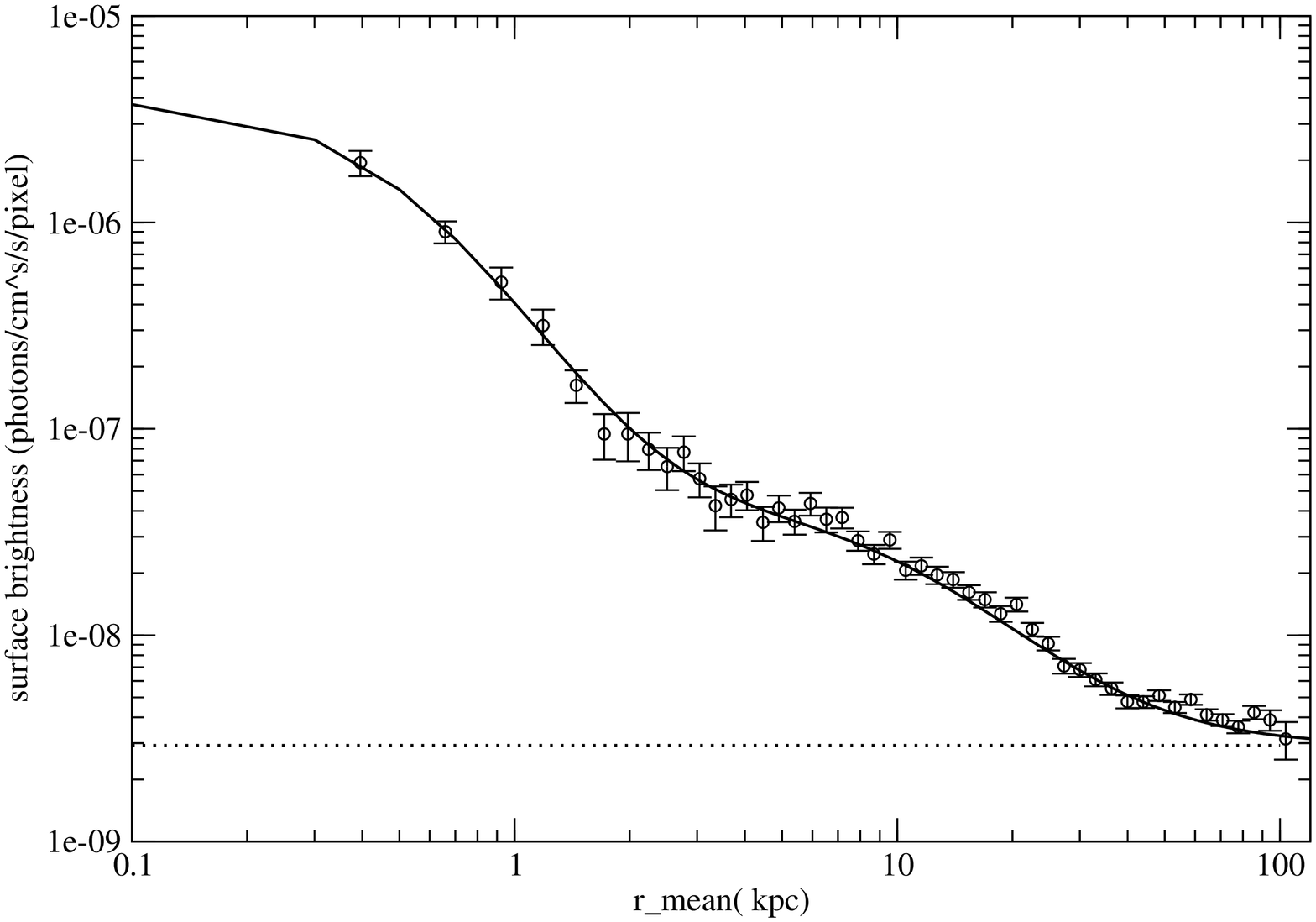}
\includegraphics[height=3.0in,width=2.37in,angle=270]{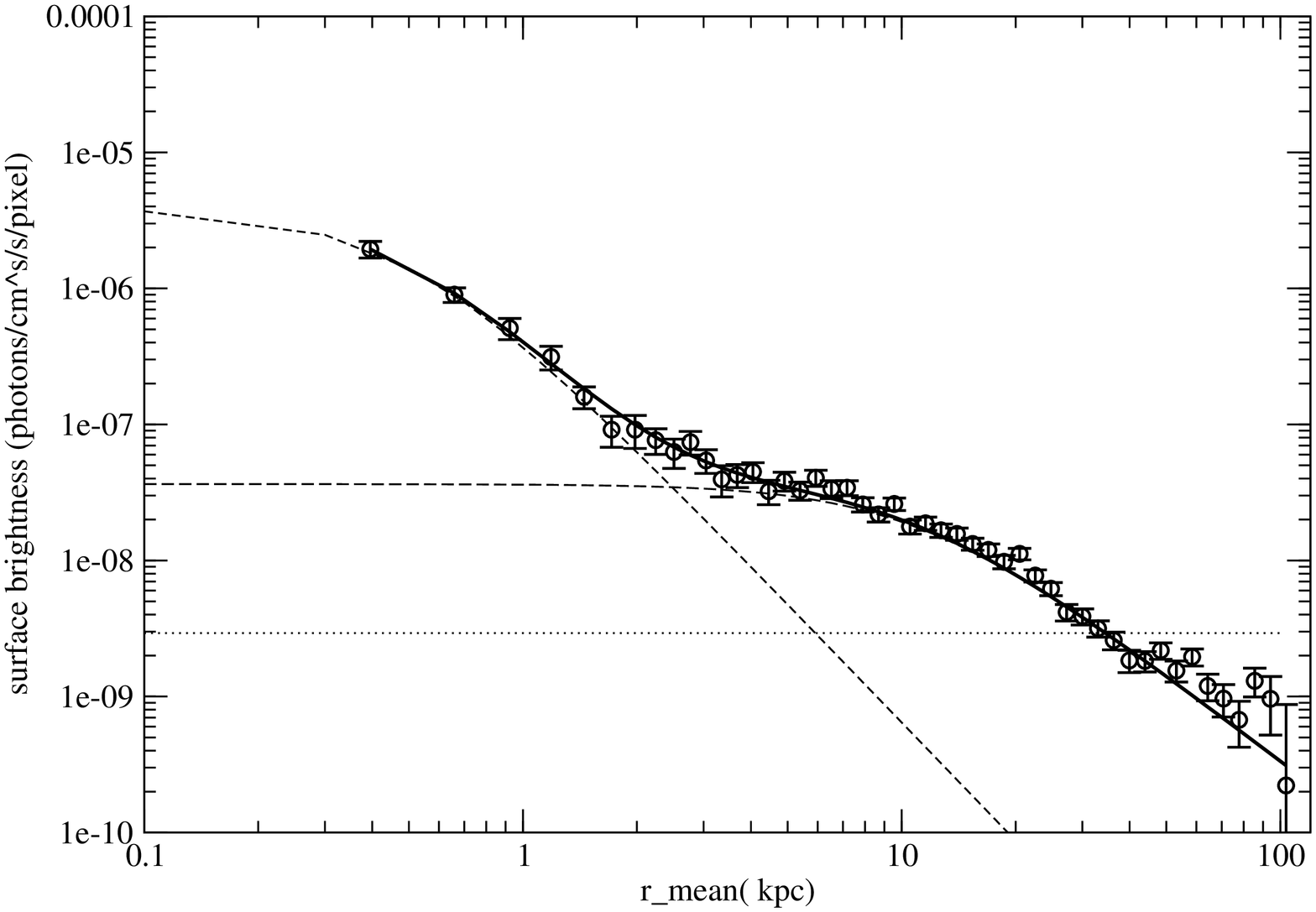}
\caption{\footnotesize{({\it upper}) Azimuthally averaged $0.5-2$\,keV 
 radial surface brightness profile of NGC\,6868 constructed using
 concentric circular annuli, centered on NGC\,6868's nucleus, with
 logarithmically increasing width. The solid line denotes the sum of
 the double $\beta$-model shown in the lower panel and a constant 
($7.5 \times 10^{-10}$\sbasunit, dotted line)
residual soft Galactic X-ray background.
({\it lower}) The azimuthally averaged surface brightness profile of
NGC\,6868 from the upper panel after subtraction of the soft Galactic
X-ray component (dotted line). The solid line is the sum of  
two $\beta$-models (shown as dashed lines), an inner model with 
index $\beta = 0.65$ and core radius $r_c=0.492$\,kpc and an outer 
model with $\beta = 0.52$ and $r_c=11$\,kpc. The dotted line denotes
the constant soft background level and thus when signal-to-noise 
$S/N = 1$. 
  }}
\label{fig:n6868meanprof}
\end{center}
\end{figure}

\subsection{NGC\,6868}
\label{sec:meanprof}

In Figure \ref{fig:n6868meanprof} (open circles, upper panel) we show the mean
radial surface brightness profile for NGC\,6868 constructed  using 
concentric circular annuli of logarithmically increasing width, 
centered on NGC\,6868's nucleus. The profile flattens at large radii 
indicative of a residual soft X-ray background component in the data. 
ROSAT All Sky Image (RASS) maps confirm that NGC\,6868 and NGC\,6861 lie in a
region of enhanced soft Galactic X-ray emission. Since this soft
Galactic X-ray background is expected to be constant over the field of
view, we estimate this background using the mean $0.5-2$\,keV 
emission in the S2 CCD from OBSID 3190, the most distant 
($r_{\rm mean} \sim 180$\,kpc) source free region from NGC\,6868,
where we expect the soft Galactic background to dominate.
We find the soft Galactic background level to be 
$7.5 \times 10^{-10}$\sbasunit
and denote this level as the horizontal
dotted line in both panels of Figure \ref{fig:n6868meanprof}. 
In the lower panel of Figure \ref{fig:n6868meanprof} we show the
surface brightness profile for NGC\,6868 in the $0.5-2$\,keV energy
 after subtraction of this residual soft Galactic background. 
The mean surface brightness profile is well
parameterized by the sum of two $\beta$-models, an inner $\beta$-model
with $\beta = 0.65$ and core radius $r_{c}=0.49$\,kpc, and an outer 
$\beta$-model  with $\beta = 0.52$ and  $r_{c}=11$\,kpc. Note that
beyond $r \sim 35$\,kpc the fit becomes highly uncertain since the
soft Galactic background dominates the faint group emission 
in this energy band. 

\begin{figure}[t]
\begin{center}
\includegraphics[height=2.29in,width=3in,angle=0]{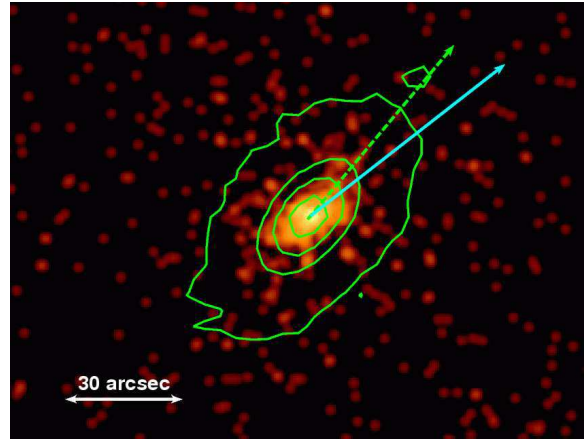}
\includegraphics[height=3in,width=2.33in,angle=270]{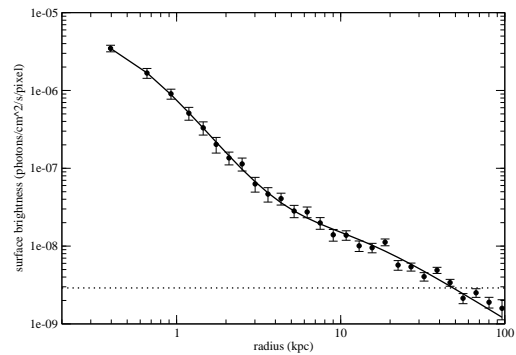}
\caption{\footnotesize{ ({\it upper}) $0.5-2$\,keV {\it Chandra} X-ray 
image of the bright inner emission region of NGC\,6861 with 2MASS
K-band contours superposed.  Resolved point sources have been
excluded and the image has been smoothed with a
$\sigma = 0\farcs984$ Gaussian kernel. K band contour levels are
$560$, $580$, $610$, $720$ data units, respectively. The dashed arrow
shows the 2MASS K band position angle ($-40^\circ $ counter-clockwise 
from North), while the solid arrow is the position angle 
($-52^\circ \pm 7^\circ$) determined from the elliptical $\beta$-model fit 
($\beta=0.62$,$r_c=0.62$\,kpc, $e=0.4$) to the
$0.5-2$\,keV X-ray surface brightness profile within $7.3$\,kpc
($54''$) of NGC\,6861. $1 {\rm pixel} = 0\farcs492 \times 0\farcs492$. 
({\it lower}) Azimuthally averaged surface
brightness profile of NGC\,6861 to large radii. Solid line denotes the 
double spherical $\beta$-model fit where $\beta=0.61 (0.38)$ and  
$r_c= 0.5 (11.9)$\,kpc  for the inner (outer) $\beta$-model component, 
respectively.)  
}}
\label{fig:n6861sbfits}
\end{center}
\end{figure}
\subsection{NGC\,6861}
\label{sec:n6861meanprof}

The distribution of starlight in NGC\,6861 is highly eccentric 
(see Fig. \ref{fig:dss}). From 2MASS K band measurements, the 
ellipticity ($e = 1-b/a$) within the ``total'' K-band isophotal semi-major
radius $a = 97\farcs4$ ($13$\,kpc) is $0.4$ with position angle, 
measured counter-clockwise from north (J2000) of $140^\circ$ (NED).  
The brightest X-ray emission, shown in Figure \ref{fig:n6861sbfits}, 
is also eccentric and highly compact. 
Using the CIAO tool Sherpa to fit a single two-dimensional 
elliptical $\beta$-model plus constant background to this central 
emission  for $r \leq 54''$ ($7.3$\,kpc), we find an eccentricity 
$0.40 \pm 0.04$, consistent with that of the starlight, but with position angle
$128^\circ \pm 7^\circ$, twisted by less than $2\sigma$ relative to 
position angle measured for the stellar light. 
We find $\beta = 0.62^{+0.02}_{-0.01}$ and a core
radius $r_c = 0.62^{+0.06}_{-0.03}$\,kpc, similar to  elliptical beta model
fits to other elliptical galaxies in galaxy groups (Osmond \& Ponman 2004). 
As we increase the fit radius to include fainter emission, the
eccentricity decreases ($e = 0.19 \pm 0.05$ for $r \leq 42$\,kpc), as 
expected as we include emission from gas residing in the more  
spherical gravitational potential of the group or subgroup. 
To separate the possible subgroup emission from
the galaxy component, we fit the background-subtracted, azimuthally-
averaged surface brightness profile, shown in 
Figure \ref{fig:n6861sbfits}, with the sum of two spherically
symmetric beta models (as we did for NGC6868 in \S\ref{sec:meanprof}).
The soft Galactic background denoted by the dashed line in Figure
\ref{fig:n6861sbfits} has also been subtracted. We  
find ($\beta$,$r_{c}$) of ($0.61$, $0.5$\,kpc) for the inner 
(galactic) component and ($0.38$, $11.9$\,kpc) for the extended 
(associated subgroup) component. The extended component $\beta$ parameter for
NGC\,6861 is much shallower than that found for NGC\,6868, but still 
falls within the range of $\beta$ parameters 
($0.36 \leq \beta \leq 0.58$) found for the extended X-ray components around 
elliptical galaxies in groups in the GEMS survey (Osmond \& Ponman
2004).

\begin{figure*}[t]
\begin{center}
\includegraphics[height=3in,width=3in,angle=90]{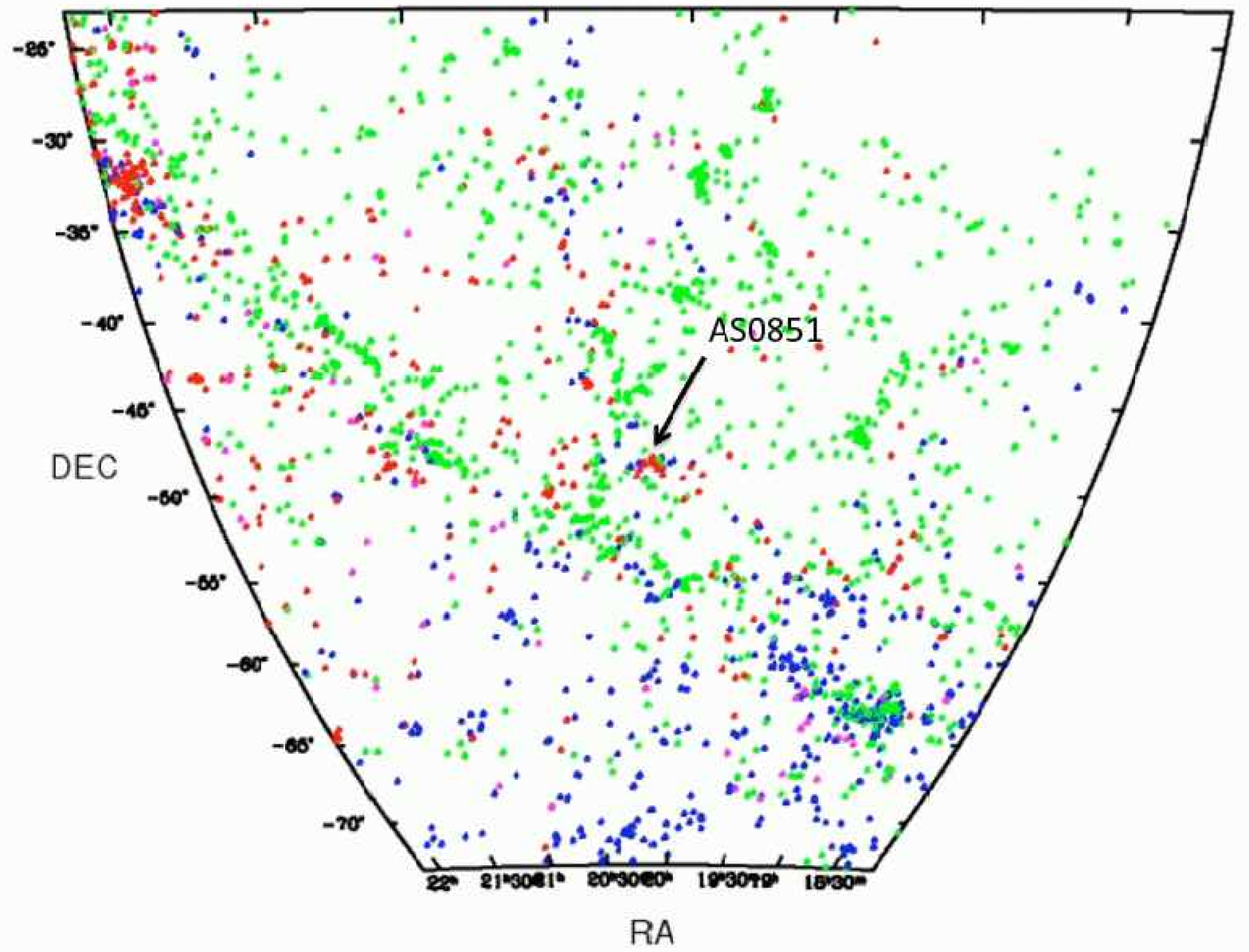}
\includegraphics[height=3in,width=3in,angle=90]{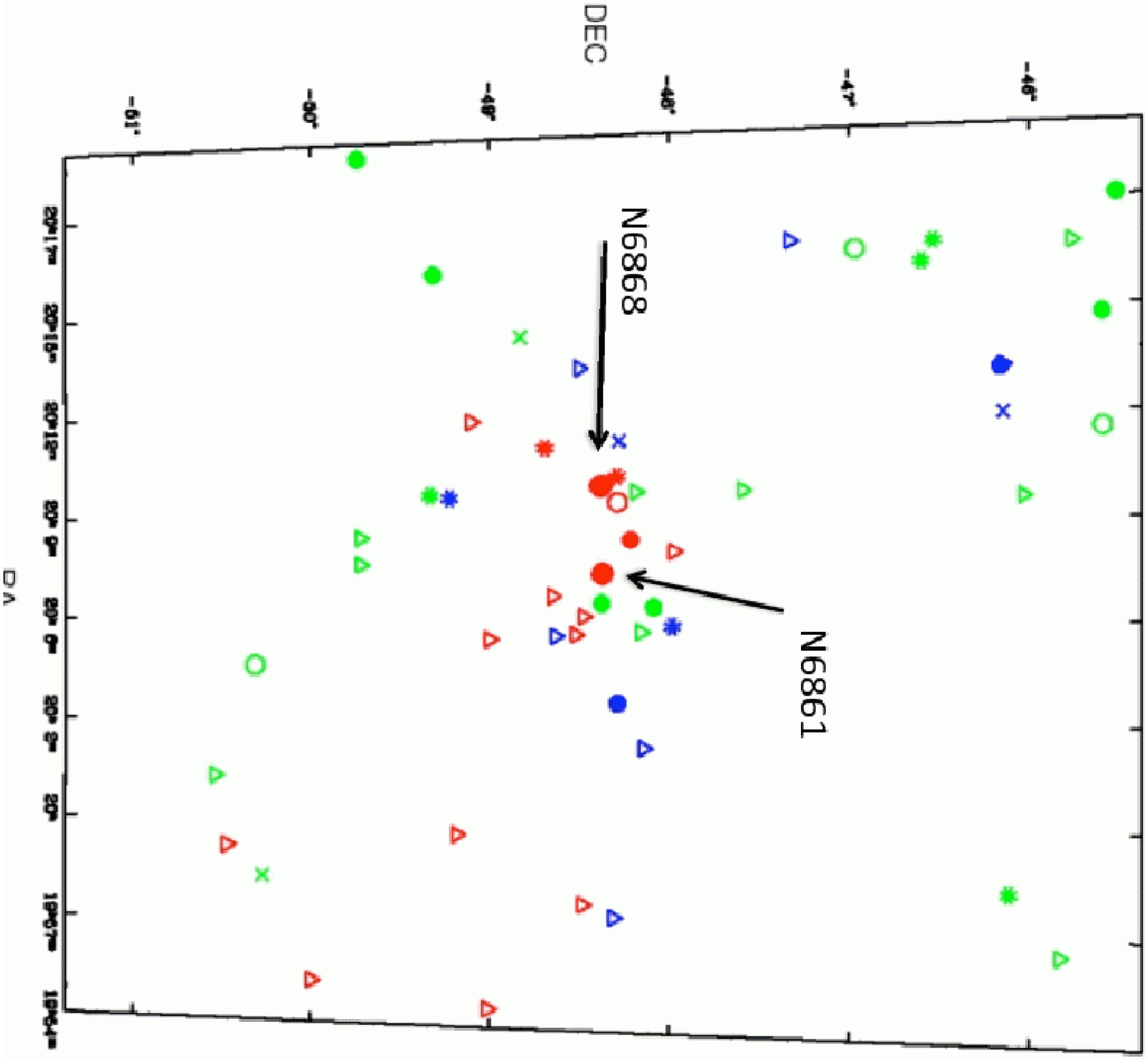}
\includegraphics[height=3in,width=2.4in,angle=270]{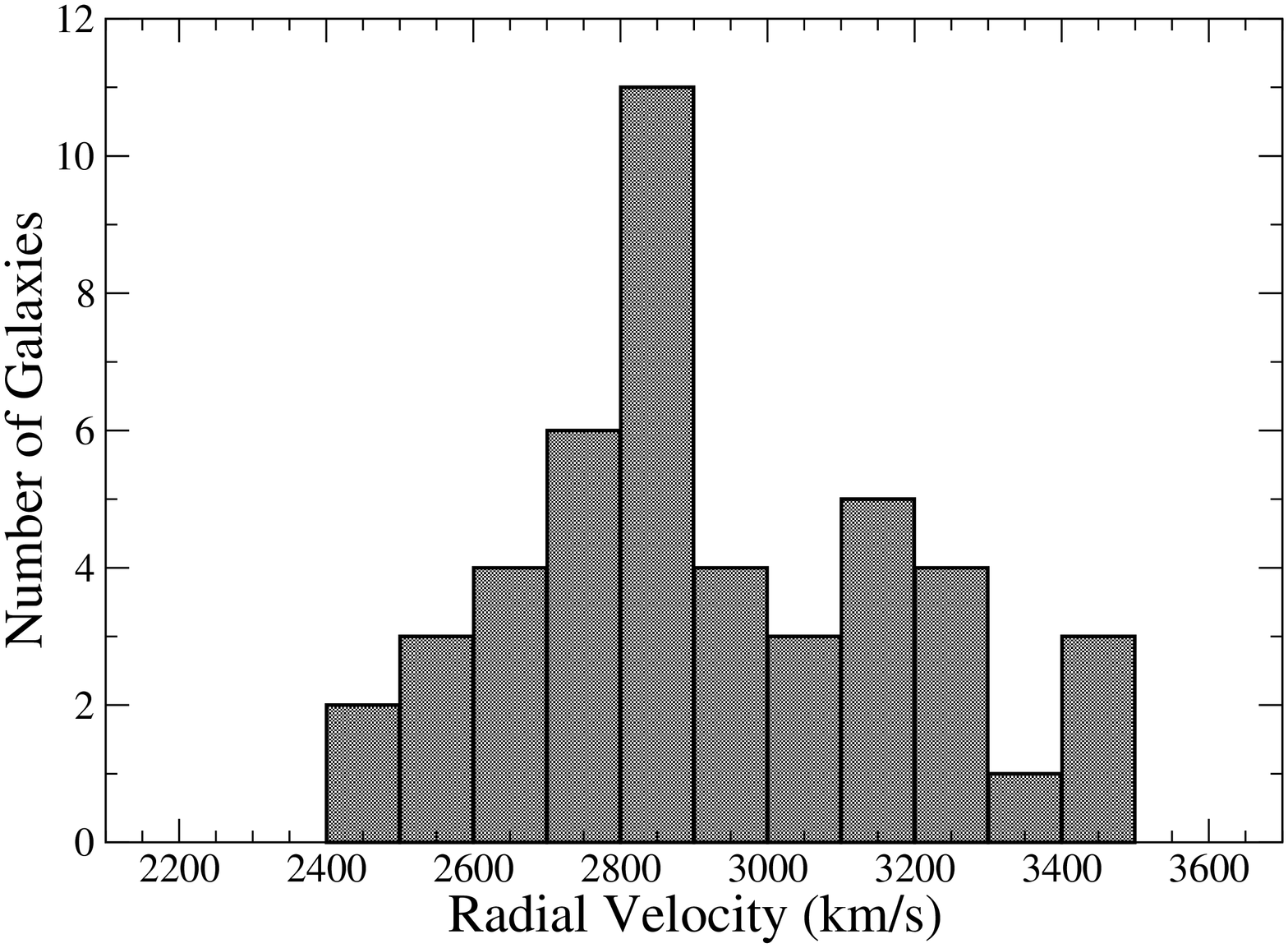}
\includegraphics[height=3in,width=2.4in,angle=270]{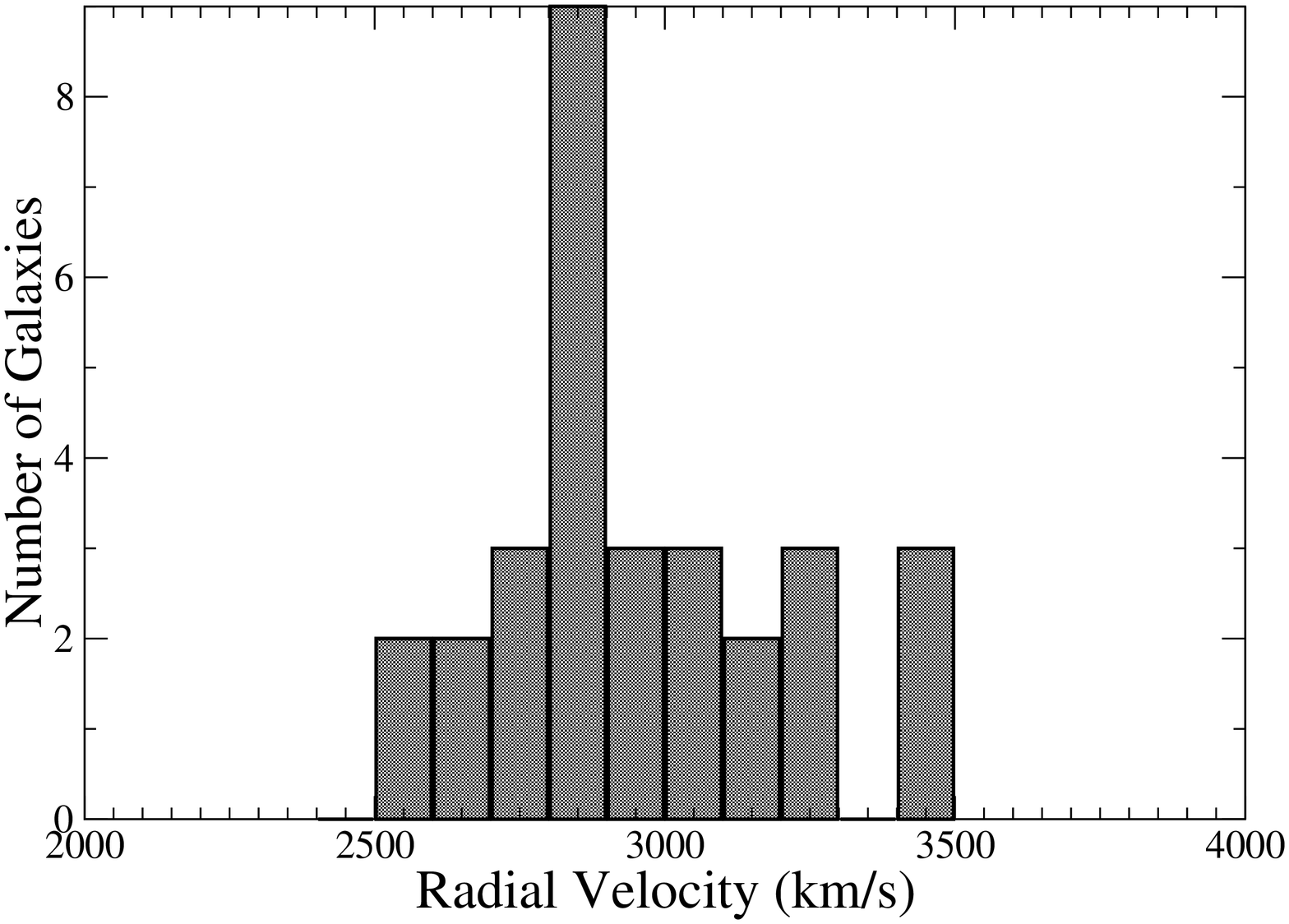}
\caption{\footnotesize{({\it upper left}) Large scale structure 
 environment of the Telescopium (AS0851) group from the CfA Redshift Survery 
 2008. Colors denote radial velocity bins of $<1500$\kms 
  (magenta), $1500 - 2999$\kms (red), $3000-4499$\,kms(blue), 
  $4500 -5999$\kms (green). The arrow denotes the position of the 
  AS0851 galaxy group. Shapes indicate galaxy morphology based on T
  type: solid circles (ellipticals, ${\rm T} = -7$ to$ -3$); open
  circle (spheroidals, ${\rm T} = -2$ to $0$); asterisk (
  early type spirals Sa-c, ${\rm T} = 1$ to $5$); crosses 
  (late type spirals and irregulars Sd - Im, ${\rm T} = 6$ to $10$);
   triangles (unclassified) and increasing symbol size denotes decreasing
   magnitude. Coordinates are J2000.0. 
{(\it upper right)} Galaxy distribution
  within $2^\circ$ of the brightest group galaxy NGC\,6868 with the
  same symbols as in the upper left panel.
 ({\it lower left}) Radial velocity histogram of 
  all galaxies with radial velocities $2000 < v_r < 4000$\kms within 
  $4^\circ$ ($2$\,Mpc) of NGC\,6868. ({\it lower right}) Radial velocity
histogram of all galaxies with radial velocities $2000 < v_r <
4000$\kms within a $1$\,Mpc ($124'$) radius of NGC\,6868. }
}
\label{fig:lss}
\end{center}
\end{figure*}

\subsection{Group Structure}
\label{sec:merger}

The fact that the surface brightness distribution in and around 
NGC\,6868 and NGC\,6861, the two brightest galaxies in AS0851, are
both described by double $\beta$-models suggests that these
galaxies may be the dominant galaxies in two subgroups that are merging.
To investigate whether there is evidence for subgroup substructure in 
the radial velocity distribution of galaxies in the AS0851 group, we
identify all galaxies with measured radial velocities 
$2000 < v_r < 4000$\kms within $4^\circ$ ($2$\,Mpc) of the 
brightest group galaxy NGC\,6868 (NED; Huchra \& Geller, CfA Redshift
Survey 2008).\footnote{http://www.cfa.harvard.edu/$\sim$huchra/zcat/.} We 
find $46$ galaxies, with radial velocities closely grouped between 
$2400 < v_r < 3700$\kms and well removed in redshift from prominent,
background large scale filaments at $v_r \sim 5000$\kms that appear nearby 
in projection (see the upper left panel of Fig. \ref{fig:lss}). 
Using the method of Osmond \& Ponman (2004), we find a velocity
dispersion for these $46$ galaxies of $274 \pm 29$\kms. 
While this velocity dispersion is typical of that
expected for a cool $1$\,keV group, the spatial extent is much larger, i.e. 
$r_{500}$ is expected to be only $\sim 0.6$\,Mpc for a $1$\,keV group
(Osmond \& Ponman 2004), which may argue in favor of the merger of
two cool galaxy subgroups. However, as seen in 
the lower left panel of Figure \ref{fig:lss}, there is little evidence 
for substructure in radial velocity space even on these larger scales.
$65\%$ of the galaxies in
this sample ($30$ of $46$) are clustered within $1$\,Mpc of NGC\,6868
or NGC\,6861 and $52\%$ (24 galaxies) are within $r_{500}$ ($0.6$\,Mpc)
of NGC\,6868. The group velocity dispersion of galaxies within 
$2$\,Mpc, $1$\,Mpc and $0.6$\,Mpc ($274 \pm 29$\kms, $263 \pm 34$\kms, 
and $251 \pm 37$\,kms, respectively) all agree within errors. 
NGC\,6868 and NGC\,6861 have radial velocities separated by only 
$35$\kms and both radial velocities lie close 
to the median radial velocity of the galaxy sample as a whole
 ($\sim 2853$\kms). Thus the merger of the NGC\,6868 and
 NGC\,6861 subgroups is predominantly in the
 plane of the sky, and it is unclear which subgroup, if either, 
 may be at the center of the AS0851 group potential. 

\section{Evidence for Interactions}
\label{sec:interact}

Galaxy interactions and the dynamical evolution 
of the surrounding IGM 
are revealed through asymmetries in the density and temperature of the gas.  
When the relative motion between the galaxy or subgroup gas halo and any 
ambient medium lies close to the plane of the sky, we expect ram 
pressure acting on the infalling body to 
produce a sharp X-ray surface brightness discontinuity along the 
leading `edge' of the infall trajectory, associated with a cold front, 
as well as a trailing tail or wake of stripped galaxy or subgroup gas
(see, e.g. Forman \etal 1979, Rangarajan \etal 1995, 
Randall \etal 2008 for M86; Scharf \etal 2004, Machacek \etal 2005a for
NGC\,1404; Machacek \etal 2006 for NGC\,4552). The surface brightness 
morphology of the gas around NGC\,6868 suggests motion. In particular, 
the surface brightness distribution around NGC\,6868, 
shown in Figure \ref{fig:chandramosaic}, appears flattened to the 
north, with a sharp discontinuity (edge) to the north and west. 
X-ray emission from NGC\,6868 is more extended (tail-like) 
to the south and east. Other asymmetries in the surface brightness
distribution include a bright clump of emission   
$ \sim 1'$ ($8$\,kpc) to the east of the nucleus, and a roughly spherical 
deficit in surface brightness in the IGM to the northwest of
NGC\,6868, as well as a partial rim-like feature in the ISM gas 
to the southwest. These latter features may be evidence for X-ray 
cavities inflated by prior AGN activity.

\begin{figure}[t]
\begin{center}
\includegraphics[height=2.29in,width=3in]{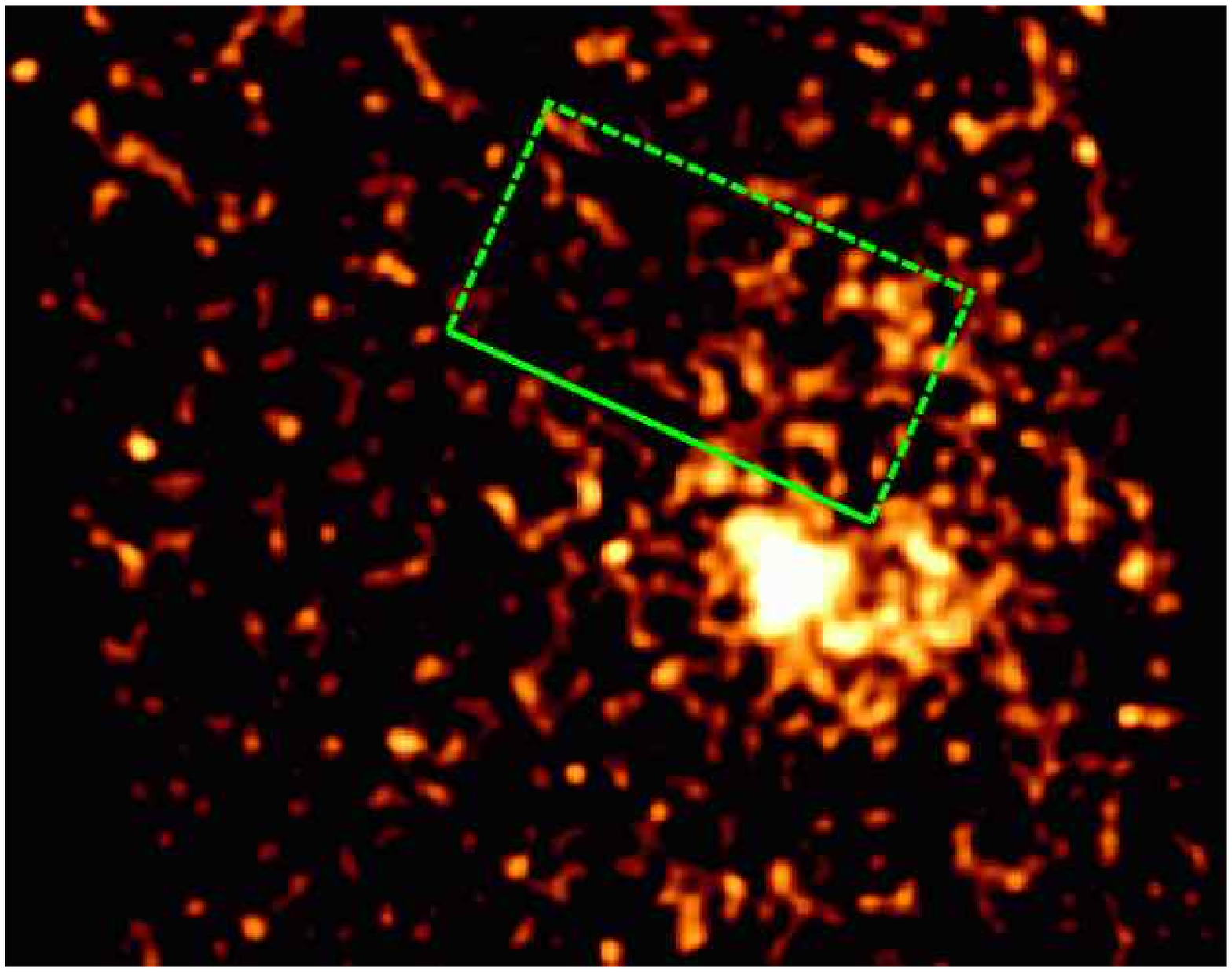}
\includegraphics[height=2.29in,width=3in]{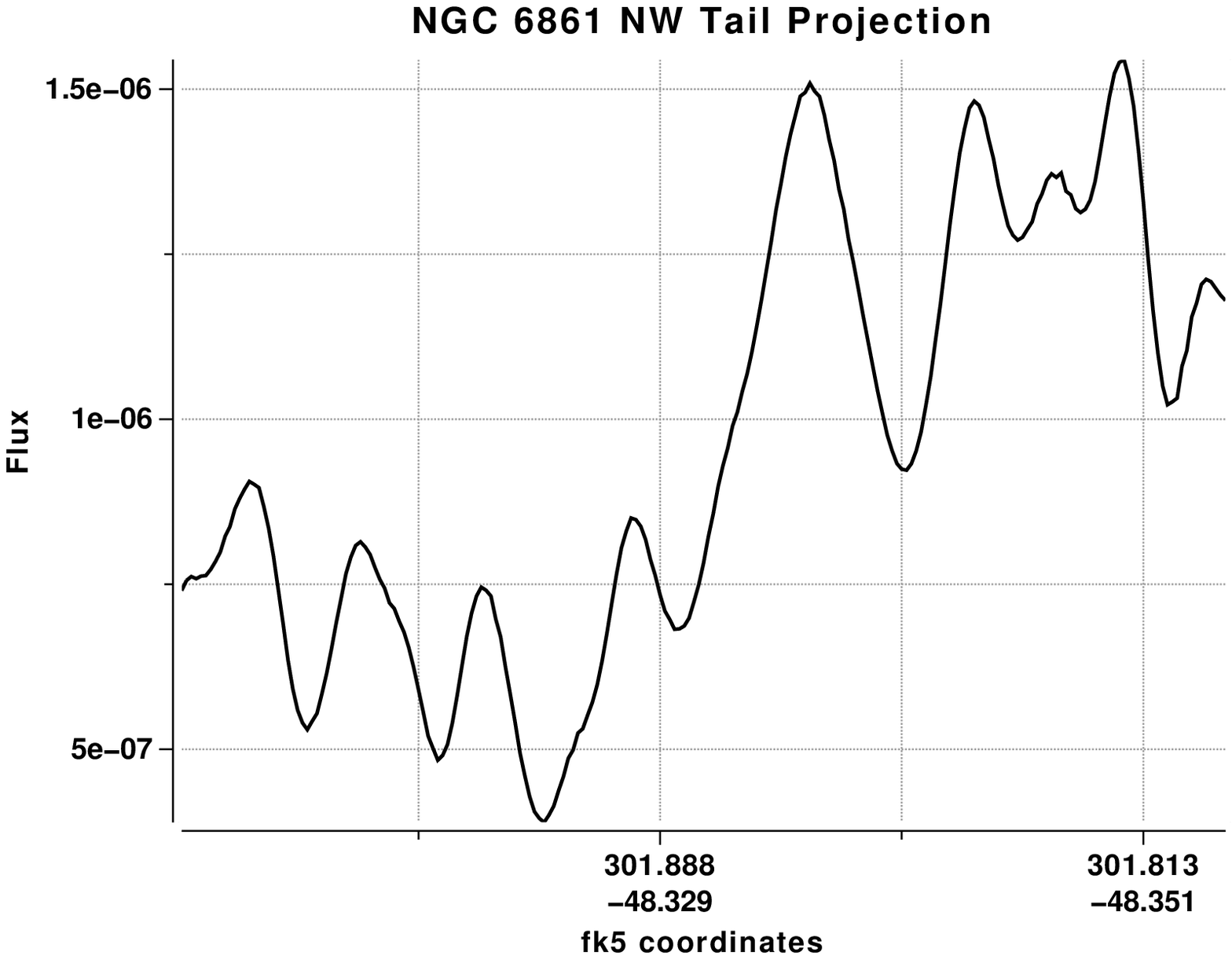}
\caption{\footnotesize{ ({\it upper}) Background subtracted, exposure 
corrected  $0.5-2$\,keV {\it Chandra} X-ray image of NGC6861 showing a 
plume of emission to northeast, a bright tail to the west, and 
a fainter tail-like feature extending to the northwest. 
The image has been smoothed with an $8''$ gaussian kernel. 
({\it lower}) Projection of surface brightness in the rectangular region 
in the upper panel, showing a factor $\sim 2$ enhancement in surface 
brightness in the northwestern tail over that in the adjacent IGM.
}}
\label{fig:nwtail}
\end{center}
\end{figure}

Although the brightest X-ray emission in NGC\,6861 follows the optical
isophotes, it is extremely compact for a massive galaxy 
and the asymmetries in the X-ray
surface brightness revealed at larger radii 
are striking (see Figs. \ref{fig:chandramosaic} and
\ref{fig:nwtail}). 
An approximately  $3'$($24$\,kpc) long X-ray tail (W Tail) 
extends due west of NGC\,6861.
A plume of emission rises $70''$ ($9.4$)\,kpc to the
northeast, perpendicular to the major axis of the galaxy,  and  
a tail of X-ray emission trails NGC\,6861 to the northwest (NW tail). 

Several scenarios offer possible explanations for the 
distorted surface brightness 
morphologies observed in and around NGC\,6868 and NGC\,6861.  
First, NGC\,6868, NGC\,6861 or both galaxies may be moving 
with respect to the group IGM, such that the extended tail-like
features may be the result of ram
pressure (Gunn \& Gott 1972) and/or turbulent-viscous  
(Nulsen 1982; Quilis \etal 2000) stripping of galaxy gas, or
gravitational focusing of the group IGM (Bondi \& Hoyle 1944; Bondi
1952; Ruderman \& Spiegel 1971; Machacek \etal 2005b). 
Second, we may be witnessing a more evolved merger of two massive
subgroups whose  dominant galaxies are NGC\,6868 and NGC\,6861, respectively, 
such that much of the NGC\,6861's outer stellar halo and  
 subgroup gas has been tidally stripped after  
multiple passes through the core of AS0851 
(Byrd \& Valtonen 1990; Moore \etal 1996) and the 
group IGM set in motion and mixed as a result of these gravitational 
encounters (sloshing, see, e.g., Markevitch \etal 2001; Mazzotta \etal 2003; 
Tittley \& Henriksen 2005; Ascasibar \& Markevitch 2006).  
Third, the disturbed morphologies may partly be the 
result of energetic outbursts from the active nuclei at the centers of the
dominant group galaxies, possibly triggered by merger activity (see, e.g.
David \etal 2009 for a study of the NGC\,5044 group).
The dynamical processes in these scenarios may act individually 
or together to effect the transformation of these galaxies and 
their environment.

\subsection{NGC\,6868's Edge and Tail: the Case for Infall}
\label{sec:n6868edge}

\begin{figure}[t]
\begin{center}
\includegraphics[height=2.76in,width=3.0in,angle=0]{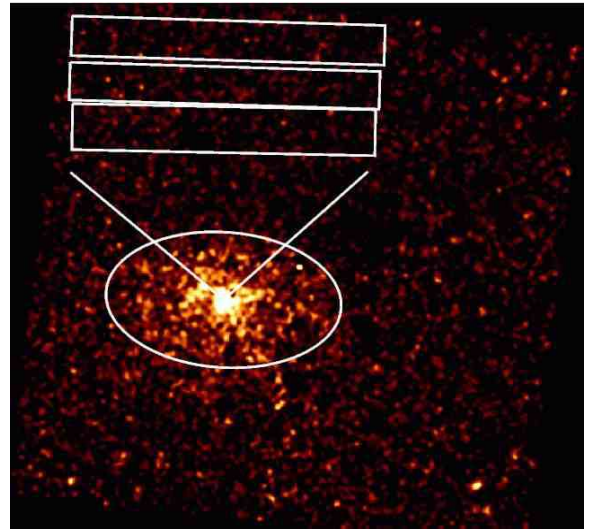}
\caption{\footnotesize{$0.5-2$\,keV blank-sky background 
subtracted, exposure corrected {\it  Chandra} image of NGC\,6868 with
the angular sector ($41.5^\circ$ to $164.5^\circ$), bounding ellipse 
centered on NGC\,6868's nucleus with (major, minor) axes and 
position angle of (228\farcs3, 131\farcs9) and
$96^\circ$, respectively. Outer box regions, used to construct 
the surface brightness profile at large radii, are superposed.}}
\label{fig:edgereg}
\end{center}
\end{figure}
\subsubsection{Modeling the Edge}

\begin{figure}[t]
\begin{center}
\includegraphics[height=3.0in,width=2.37in,angle=270]{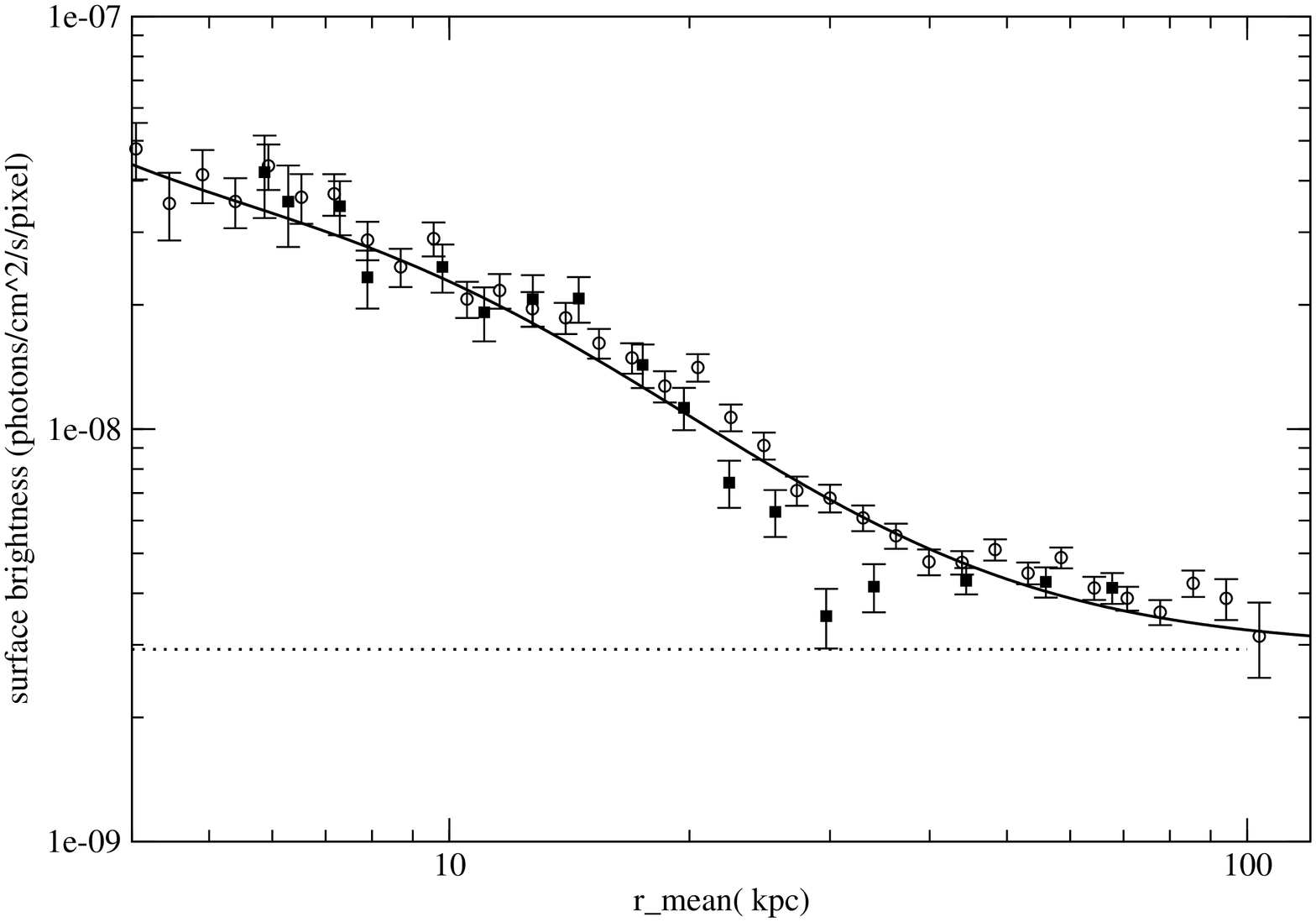}
\includegraphics[height=3.0in,width=2.37in,angle=270]{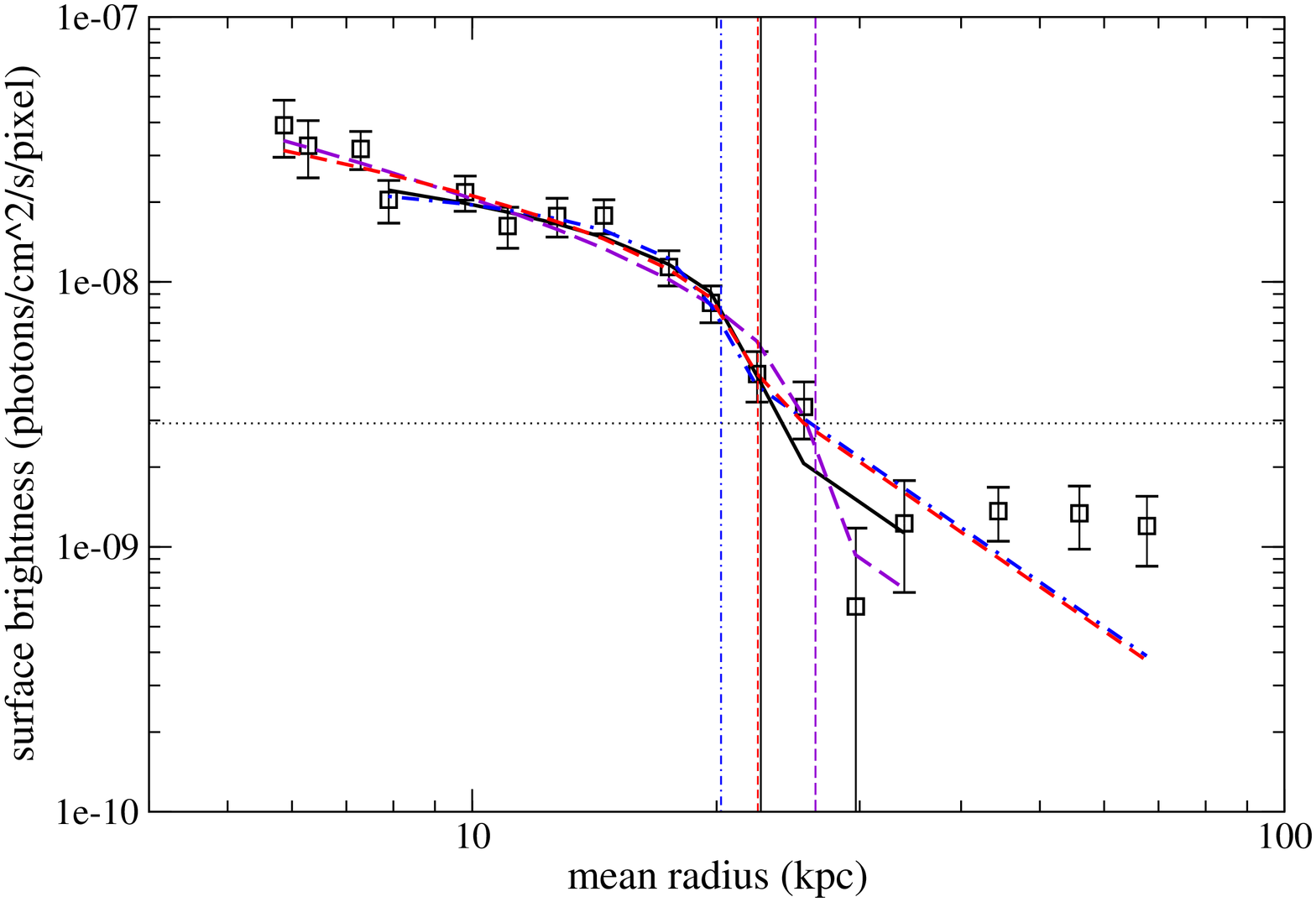}
\caption{{\footnotesize ({\it upper}) $0.5-2$\,keV surface brightness
  profile across the northern edge of NGC\,6868 (filled squares), 
using the regions shown in \protect\ref{fig:edgereg}, superposed 
on the azimuthally averaged $0.5-2$\,keV surface brightness 
profile from the upper panel of \protect\ref{fig:n6868meanprof} (open circles).
The lines are defined as in Fig. \protect\ref{fig:n6868meanprof}.
({\it lower}) Model fits to the northern surface brightness edge in
NGC\,6868 in the sector shown in Fig. \protect\ref{fig:edgereg}. 
Open squares denote the surface brightness profile after subtraction
of the soft Galactic background. Curved lines denote the density model
fits to the background-subtracted surface brightness profile across
the edge over the radius range $7.9-34$\,kpc (solid black), 
$7.9-68$\,kpc (dot-dash blue), $5.8-35$\,kpc (long dash violet), 
and $5.8-68$\,kpc (short dash red) listed in 
Table \protect\ref{tab:n6868edgefits}. The corresponding vertical
lines denote the edge position in each case and the dotted horizontal 
line shows where $S/N = 1$.
  }}
\label{fig:n6868edgefit}
\end{center}
\end{figure}

The sharp X-ray surface brightness edge to the north of NGC\,6868 and 
trailing tail to the south (see Fig. \ref{fig:chandramosaic}), 
suggest  that NGC\,6868 may be moving with respect to the group IGM. 
In Figures \ref{fig:edgereg} and \ref{fig:n6868edgefit} we 
investigate the northern edge-like feature.
Following Vikhlinin \etal (2001), 
we construct the $0.5 - 2$\,keV surface brightness
profile for NGC\,6868 across the edge using elliptical annuli, 
centered on NGC\,6868's nucleus and 
constrained to lie in an angular sector extending from $41.5^\circ$
to $164.5^\circ$, to avoid the northwestern cavity. The elliptical
annuli are chosen concentric to a bounding ellipse that traces the surface
brightness discontinuity in the sector of interest 
(see Fig. \ref{fig:edgereg}). The width of each 
elliptical annulus increases (decreases) with constant logarithmic step
size as one moves outward (inward) from the bounding ellipse.   
Beyond $50$\,kpc, the elliptical annuli are no longer fully 
contained within the ACIS-I field-of-view. We define three narrow 
rectangular regions (also shown in Fig. \ref{fig:edgereg}) 
to extend the surface brightness profile to larger radii. 

The resulting profile is shown in Figure \ref{fig:n6868edgefit}. 
 In the upper panel of Figure \ref{fig:n6868edgefit} we
compare the surface brightness profile across the northern edge (filled
squares) to the azimuthally averaged profile for the galaxy (open
circles). We see a sharp, statistically significant, drop in surface 
brightness between  $20 \lesssim r \lesssim 35$\,kpc, consistent with
the presence of an edge. 
To model the X-ray surface brightness profile across the edge, 
we integrate the square of the electron density $n_e$ times the
corresponding X-ray emissivity $\Lambda$ along the line of sight,
assuming that the electron density on either side of the discontinuity
is well described by spherical power law models of the form:
\begin{equation}
\begin{split}
n_e = n_1 \big (\frac{r}{r_{\rm edge}}\big )^{-\alpha 1},  \qquad r
< r_{\rm edge} \\
n_e = n_2 \big (\frac{r}{r_{\rm edge}}\big )^{-\alpha 2},  \qquad r
\geq r_{\rm edge} 
\end{split}
\label{eq:densemod}
\end{equation}
where $r_{\rm edge}$ is the location of the edge. The observed  
surface brightness discontinuity is then given by 
\begin{equation}
J^2 = \Big (\frac{\Lambda_1 n_1^2}{\Lambda_2 n_2^2} \Big ),
\label{eq:jump}
\end{equation} 
where $\Lambda_i$ and $n_i$ are the X-ray emissivity and electron
density for gas inside ($i=1$) and outside ($i=2$) the edge. 
We fit the surface brightness profile across the edge using a multi-variate
chi-square minimization scheme allowing the edge position  
$r_{\rm edge}$, electron density power law index $\alpha_1$ inside 
the edge, and the size of the discontinuity (jump) $J$ to vary.  
Since the signal-to-noise drops below $1$ just outside the edge (see 
Fig. \ref{fig:n6868edgefit}), the powerlaw index $\alpha_2$ for the IGM
gas density is not well determined by these data. We fix $\alpha_2=
-1.56$ from the parameterization of the mean surface profile at 
large radii. The results of our model fits are listed in 
Table \ref{tab:n6868edgefits} and plotted in the lower panel of 
Figure \ref{fig:n6868edgefit}, where
we have varied the fit range to probe the sensitivity of the edge
position and jump discontinuity to the low S/N data.  We find 
$r_{\rm edge} \sim 23 \pm 3$\,kpc and jump discontinuity 
$J \sim  2.2^{+0.4}_{-0.5}$.  

To complete the density model, we compute the X-ray emissivity 
$\Lambda$ in the $0.5-2$\,keV energy band from APEC thermal plasma 
models for the X-ray emission from 
elliptical annular sectors, listed in Table
\ref{tab:n6868edgespecreg}, chosen concentric to the edge bounding 
ellipse (see Fig. \ref{fig:edgereg}) inside 
(NEdge$_{-1}$) and outside (NEdge$_{+0}$,NEdge$_{+1}$) the edge. The results 
of these spectral fits are given in Table \ref{tab:n6868edgespec}. Although 
the uncertainties are large, the temperature of the 
IGM outside the edge to the north is higher
($0.73^{+0.07}_{-0.09}$\,keV) than inside the edge 
($\sim 0.64 \pm 0.08$\,keV), suggesting that the edge is a cold front. 
For temperatures $\lesssim 1$\,keV, the Fe L complex line emission 
dominates the thermal continuum in the $0.5-2$\,keV energy band, 
such that the X-ray emissivity depends sensitively on the abundance of the 
gas, and can vary by a factor $\sim 5$ as the abundance varies 
from $0.1$ to $1\,\Zs$. This introduces an uncertainty in the inferred density 
ratio across the edge by a factor $\sim 2$. Although the abundances are 
poorly constrained by these data, the spectral models in  
Table \ref{tab:n6868edgespec} suggest that low 
abundances ($A \lesssim 0.3\,\Zs$) are favored in the group IGM north of 
the edge, consistent with abundances measured in other cool groups and in 
the outskirts of clusters (Rasmussen \& Ponman 2007; Finoquenov \etal
2000; De Grandi \etal 2004). The spectral models 
in Table \ref{tab:n6868edgespec} also suggest that higher abundances 
$A \sim 0.5-1\,\Zs$ are favored inside the edge closer to the center of 
NGC\,6868, consistent with abundances measured in other dominant 
elliptical galaxies (see, e.g. Finoguenov \& Jones 2001; 
Matsushita \etal 2003). 

For simplicity we define a pair of fiducial spectral models with
temperature $kT = 0.73$\,keV and abundance $A=0.3\,\Zs$ for 
IGM gas to the north, outside the edge, and 
$kT = 0.64$\,keV and  $A=0.5\Zs$ for galaxy gas inside the edge. 
Normalizing the density model using the IGM gas density 
from eq. \ref{eq:densemod} at $r = 34$\,kpc 
to the north,  where the contribution from the 
galaxy gas is negligible (see Fig. \ref{fig:n6868meanprof}), we find 
an electron density in the IGM 
$n_e(r=34\,{\rm kpc}) = 4.4 \times 10^{-4}$\cmc.
Combining this with the density model fits 
to the edge location and surface brightness jump from 
eq. \ref{eq:jump}, we find electron gas densities at $r=r_{edge}$ 
inside ($n_1$) and  outside ($n_2$) NGC\,6868's northern surface 
brightness edge to be $(1.2 - 1.5) \times 10^{-3}$\cmc and 
$(0.66-1.0) \times 10^{-3}$\cmc, respectively 
(see Table \ref{tab:n6868edgefits}). Note that if there is no
abundance gradient across the edge, the density ratio derived from 
eq. \ref{eq:jump} would be $23\%$ larger. 

\subsubsection{Velocity Constraints}

To constrain the relative motion of NGC\,6868 with respect to 
the ambient IGM, we assume 
uniform gas flow past a rigid spheroid, 
such that the difference between the gas 
pressure in the undisturbed IGM (free stream region)
and that at the stagnation point of the flow, where the relative 
velocity of the IGM gas and galaxy gas vanishes, is the ram pressure
due to the motion of the galaxy through the IGM (Vikhlinin \etal 2001). 
The pressure ratio $p_0/p_2$ between IGM gas at the stagnation point 
and that in the free-stream region allows us to determine the Mach
number of the flow (Landau \& Lifshitz 1959) and thus the
three-dimensional velocity of the infalling galaxy with respect to the
ambient gas. Since the stagnation 
region is small and difficult to observe, and pressure is
continuous across a cold front edge, we use the pressure in 
NGC\,6868 just inside the northern edge ($p_1$) as a proxy for the pressure at 
the stagnation point. We use the  density and spectral model fits to gas
inside and outside NGC\,6868's northern edge, given in 
Tables \ref{tab:n6868edgefits} and \ref{tab:n6868edgespec}, to
compute the range of pressure ratios ($p_1/p_2$) and Mach numbers
consistent with the fits to the surface brightness discontinuity (see 
Fig. \ref{fig:n6868edgefit}). Our results are given in 
Table \ref{tab:n6868mach}. 

For all models, the inferred 
velocity for NGC\,6868 relative to the surrounding group gas is at most 
transonic, i.e. $v \leq 475$\,kms (${\rm Mach} \leq 1.08$), 
given the speed of sound 
$c_s = 440^{+20}_{-30}$\kms for  $0.7$\,keV group gas. This is similar to 
that found in other infalling galaxies, such as NGC\,1404 in Fornax 
(Machacek \etal 2005a) and, if we assume  NGC\,6861 rather than
NGC\,6868 is at the group center, not unexpected given 
the $\sim 200$\,kpc ($\sim 0.3r_{500}$) 
separation between the two galaxies. On the other hand, the
uncertainties in the observed metal abundances are large, and, since  
the spectrum for cool ($\lesssim 1$\,keV) gas is
dominated by the Fe L complex line emission, the emissivity of X-ray
gas is highly dependent on the metal abundance of the gas. 
Solar abundances are often measured
in the centers of dominant elliptical galaxies in groups 
(for a compilation, see, e.g. Rasmussen \& Ponman 2007 and 
references therein). If we assume that solar abundances extend all the way
to the surface brightness discontinuity in NGC\,6868, the sharp jump
in surface brightness may be 
explained by the ratio of abundances (and thus emissivities) across
the edge alone. In that case, NGC\,6868 may be at rest with respect 
to the surrounding IGM and likely at the center of the group
potential. However, then stripping could not explain the origin 
of the southern tail. 

\subsubsection{NGC\,6868's Southern Tail}
\label{sec:southtail}

As suggested in Figure \ref{fig:chandramosaic} 
and confirmed in the lower panel of Figure \ref{fig:n6868cavity}, 
X-ray emission (and thus 
gas density) is higher to the south of the galaxy 
($171.5^\circ < \phi < 351.5^\circ$) than to 
the north at the same radii. If the southern tail is composed of
ram-pressure stripped galaxy gas, we would expect the metal abundance
in the southern tail to be higher, similar to abundances
inside the edge in the galaxy gas halo, than in the IGM 
to the north. 
To test whether the southern tail is  gas ram-pressured stripped 
from NGC\,6868, we fit the spectrum from an elliptical annular 
sector ($S_{+1}$) at the same mean radial distance and with the same 
radial bin width as sector NEdge$_{+1}$ to the north
(see Table \ref{tab:n6868edgespecreg}), but
constrained to lie in the tail to the south 
between $213^\circ$ to $327.5^\circ$. As shown in 
Table \ref{tab:n6868edgespec}, the best fit single APEC model for 
this southern region gives a gas temperature and abundance of 
$0.63^{+0.05}_{-0.06}$\,keV and $0.29^{+0.24}_{-0.18}\Zs$, respectively.  
While the temperature of the gas in this region is lower than that of the 
IGM to the north and comparable to the temperature of galaxy gas within 
NGC\,6868, as one might expect if the gas were ram-pressure stripped
galaxy gas, the uncertainties in the measured abundance are
too large to definitively determine the origin of gas in the southern tail.  
Deeper X-ray data are needed to distinguish between: (1) cool, 
higher abundance ($\sim 0.5\,Zs$)  ram-pressure stripped galaxy gas, 
(2) lower temperature, lower entropy subgroup gas with 
an abundance comparable to that found in the IGM to the north, or 
(3) a multiphase mixture of the two, possibly 
signaled by the low $90\%$\,CL lower bound on the abundance ($\sim
0.11\,\Zs$) from the single temperature APEC model fit (Buote  2000).  
     
\begin{figure}[t]
\begin{center}
\includegraphics[height=3in,width=2.33in,angle=270]{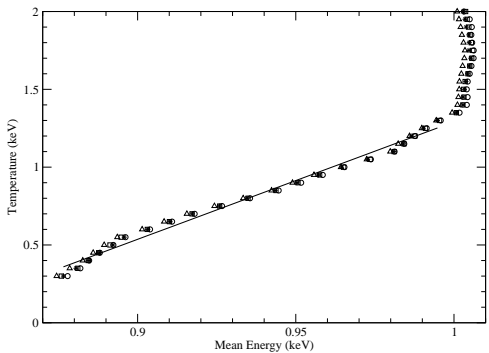}
\caption{\footnotesize{Gas Temperature as a function of mean
photon energy in the $0.7-1.2$\,keV enery band for ACIS-I CCD  
I0 (circles), I1 (squares), I2(triangles), and I3 (astericks), 
assuming a single temperature APEC model with fixed metal abundance 
$A=0.5\,\Zs$ and Galactic absorption of $3.9 \times 10^{20}$\cms. 
The solid line denotes the best linear fit ($kT = -6.258 + 7.5492\,E$) 
restricted to $0.4 \leq kT \leq 1.3$\,keV. }}
\label{fig:calib}
\end{center}
\end{figure}

\subsection{Temperature Asymmetries: the Case for Sloshing}
\label{sec:temp}

Modest density jumps and subsonic relative velocities, as observed for
NGC\,6868, are also characteristic of gas `sloshing', where the
surrounding IGM has been set in motion by the gravitational
perturbation of a prior passage of galaxies or galaxy subgroups 
past each other in the group core (Markevitch \etal 2001). 
Simulations of sloshing in
subcluster mergers show that the signatures of sloshing are most
pronounced in the temperature structure of the gas, often producing 
cool, spiral-shaped tail-like features, as higher entropy gas from 
larger radii is mixed with the nearby lower entropy galaxy and subgroup gas 
(Ascasibar \& Markevitch 2006). 
\begin{figure*}[t]
\begin{center}
\includegraphics[height=1.85in,width=3.2in]{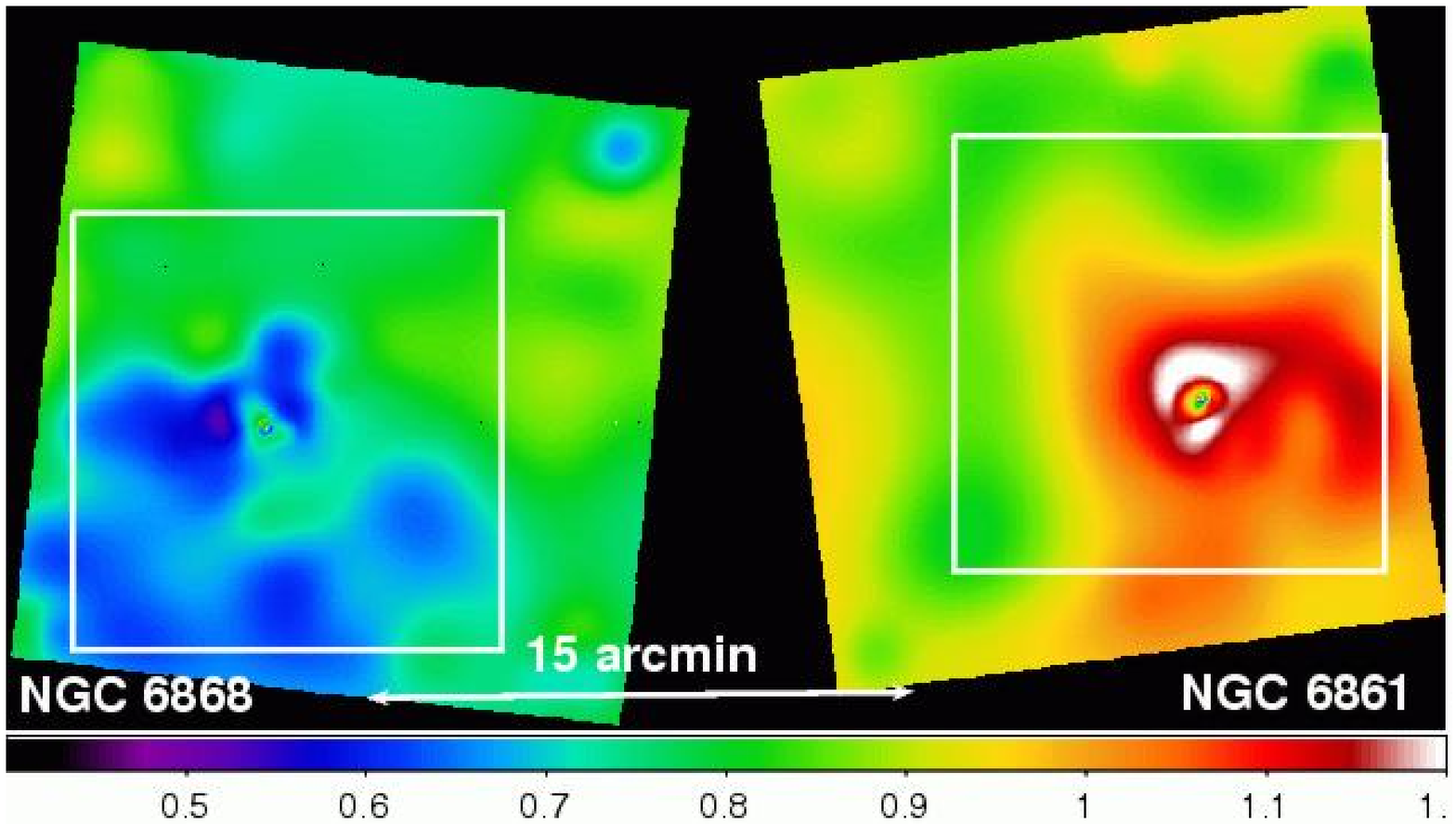} 
\includegraphics[height=1.82in,width=3.2in]{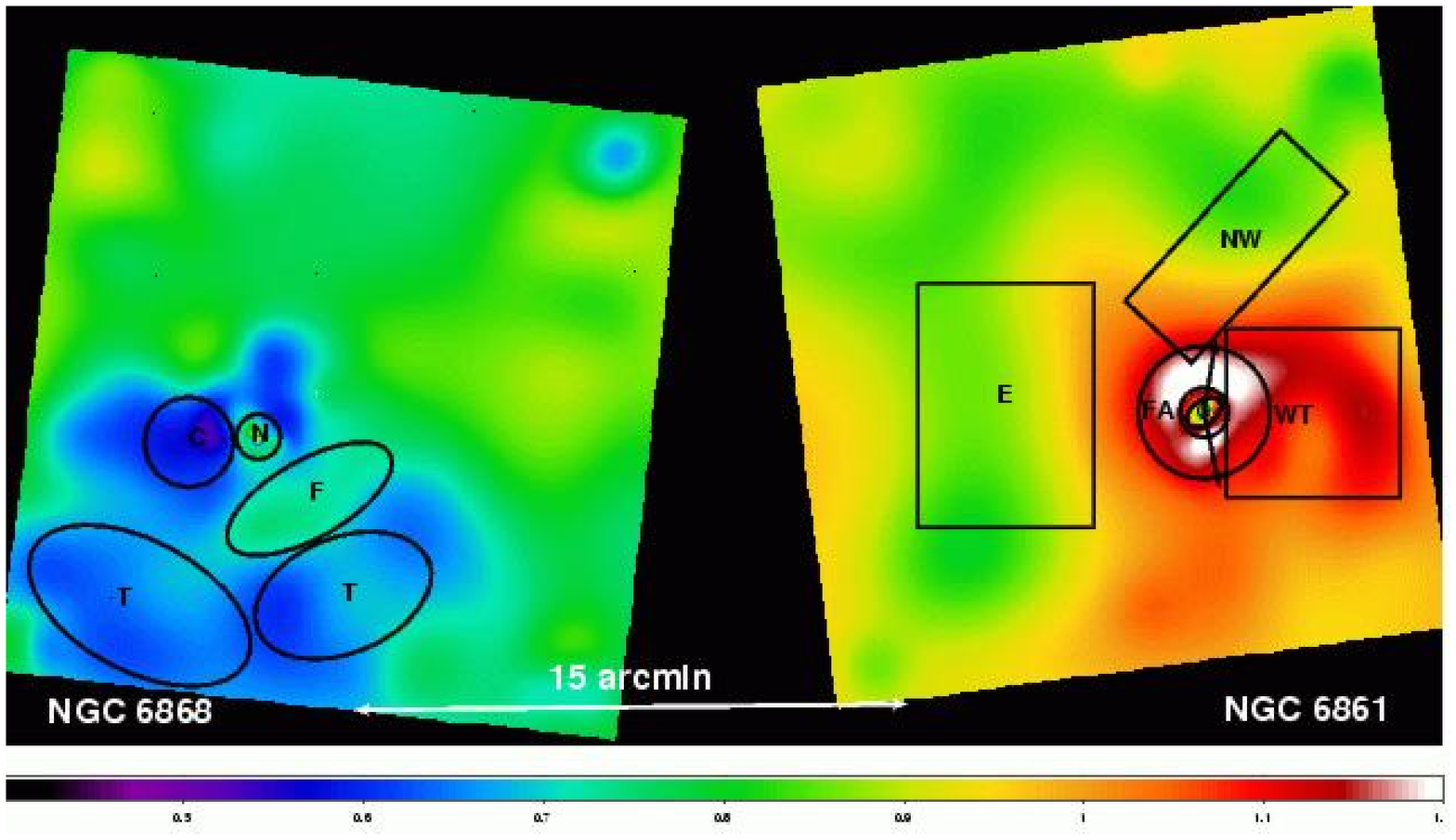}  
\includegraphics[height=1.55in,width=3.2in]{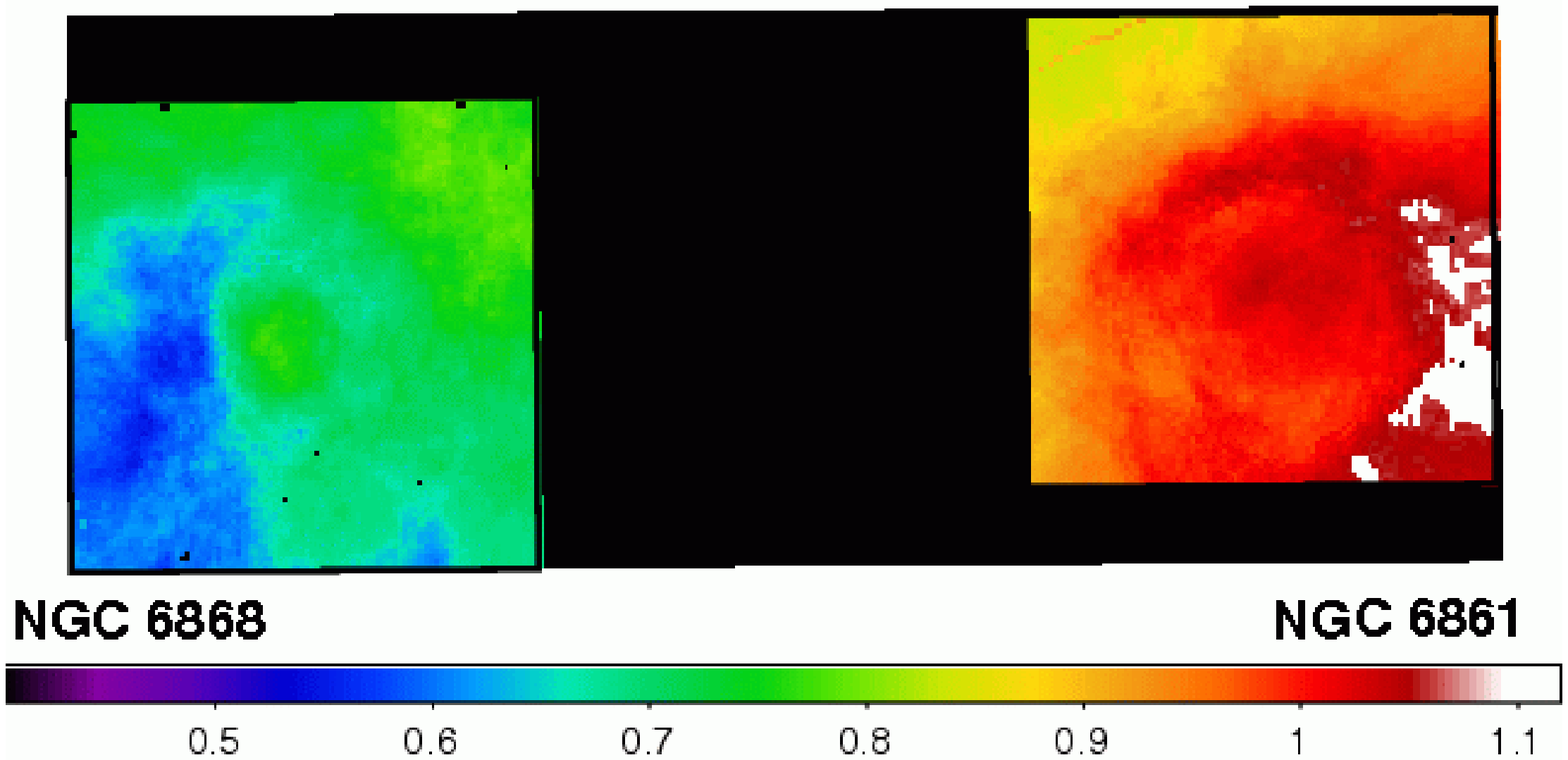}
\includegraphics[height=1.533in,width=3.2in]{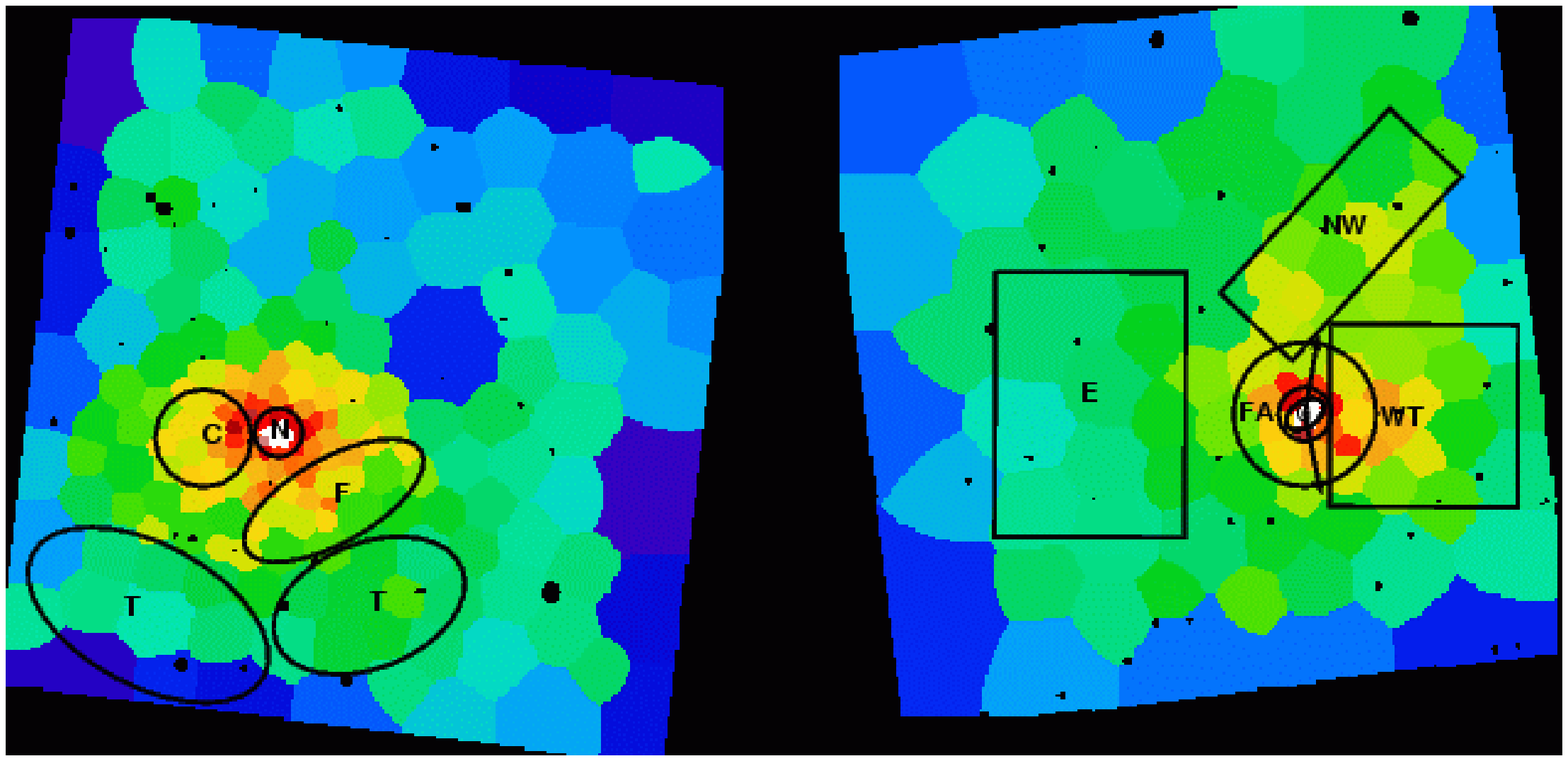}
\caption{\footnotesize{({\it upper left}) Temperature map of NGC\,6868 
and NGC\,6861 constructed from the mean photon energy in the Fe peak 
at $0.7-1.2$\,keV. Rectangular regions correspond to fields fit with 
the coarse temperature map shown below in the lower left panel 
({\it upper right}) Same temperature map  
with spectral regions from Tables \protect\ref{tab:n6868tmapspecreg} and
\protect\ref{tab:n6861specreg} superposed. N and G mark the centers of 
NGC\,6868 and NGC\,6861, respectively. 
({\it lower left}) Coarse temperature map 
constructed by explicitly fitting single
temperature APEC models to regions containing a minimum of $500$ counts. 
({\it lower right}) Spectral
regions from Tables \protect\ref{tab:n6868tmapspecreg} and
\protect\ref{tab:n6861specreg} on the adaptively binned 
$0.5-2$\,keV Chandra X-ray image of NGC\,6868 and NGC\,6861 from 
Fig. \protect\ref{fig:chandramosaic}. 
}}
\label{fig:Fepeaktmap}
\end{center}
\end{figure*}

\subsubsection{Temperature Map}
\label{sec:tmap}

Since NGC\,6868 and 
NGC\,6861 are elliptical galaxies in only a moderately massive galaxy
group, we expect temperatures for gas in the galaxies and surrounding 
group/subgroup gas to be $\lesssim 1$\,keV. For such low gas
temperatures the X-ray spectrum peaks at $\sim 1$\,keV and is
dominated by lines in the Fe L complex. Using XSpec simulations of 
single temperature APEC plasma models, we find that there is 
nearly a linear correspondence between the temperature of the X-ray 
emitting gas in the $0.4-1.3$\,keV range and the mean photon energy 
measured in the $0.7-1.2$\,keV energy band, containing the Fe L
emission (See Fig. \ref{fig:calib} and David \etal 2009). 
We exploit this correspondence to construct 
the temperature map of NGC\,6868 and NGC\,6861, shown in 
 Figure \ref{fig:Fepeaktmap}. 
After resolved point sources have been removed, we
adaptively smooth the narrow ($0.7-1.2$\,keV) band {\it Chandra}
surface brightness images for NGC\,6868 and NGC\,6861 to determine 
regions of interest. We then average the photon energy in the Fe L
peak over these regions and use the calibration shown in 
Figure \ref{fig:calib} to relate these mean energies to the gas temperature. 
To check that our results were not an artifact of the binning
procedure, we constructed a second temperature map, shown in the lower
left panel of Figure \ref{fig:Fepeaktmap},  
by explicitly fitting single temperature APEC
models to regions containing a minimum of $500$ counts (O'Sullivan
\etal 2005; Maughan \etal 2006; Randall \etal 2008). 
Although the second binning is much coarser and averages over the 
nuclear region in each galaxy, the results for the extended emission
outside the nucleus are qualitatively the same.

The temperature maps confirm that the gas temperature in NGC\,6868 
inside (south of) the surface brightness edge is lower than in the 
group IGM to the north outside the edge.
The lowest temperature gas ($\sim 0.59$\,keV) in NGC\,6868 is displaced 
$\sim 10-20$\,kpc ($1\farcm2 - 2\farcm5$) to the east of NGC\,6868's 
nucleus. There is a ring of higher temperature gas close to the
nucleus, possibly indicating nuclear activity.  
The extended gas tail to the south of NGC\,6868 is  
cool ($\sim 0.65$\,keV), and appears to spiral to the south and
southwest around a $\sim 30$\,kpc long finger of higher temperature 
emission southwest of the nucleus. 
Both temperature maps show the spiral feature in NGC\,6868's extended 
southern tail. Although the temperature of the southern tail is 
consistent with that expected for ram-pressure stripped galaxy gas 
and such tails may curve, following the orbit of the infalling galaxy
or subgroup, the tightly wound spiral pattern of cool gas wrapping 
about a higher temperature filament is more characteristic of the `sloshing' 
signatures seen in simulations of IGM gas set in motion by interactions of
galaxies or subcluster cores near the centers of clusters 
(Tittley \& Henriksen 2005; Ascasibar \& Markevitch 2006). 
Such features have been observed recently in X-ray observations of 
sub-cluster mergers and in galaxy groups (e.g see Markevitch \& 
Vikhlinin 2007 for A2142; Mazzotta \& Giacintucci 2008 for
MS1455.0+2232; Johnson \etal 2009 for A1644; Randall \etal 2009 for 
the NGC\,5098 group). 

The temperature structure of NGC\,6861 is dramatically different. In both
temperature maps, single temperature APEC models suggest that the
temperature of gas in the bright western tail is higher than in the 
surrounding IGM. Such high temperature tails have been observed in 
other galaxy groups, but are rare (see, e.g. 
Osmond \etal 2004 for the NGC\,5171 group; Machacek \etal 2005b 
for the Pavo group). The upper panels of Figure \ref{fig:Fepeaktmap} 
also show a sheath of high temperature gas $\sim 8$\,kpc ($1'$) from 
NGC\,6861's center, well outside the X-ray bright central region of 
NGC\,6861. There is no feature in the temperature maps corresponding 
to the northwestern tail observed in the X-ray surface brightness 
images (Figure \ref{fig:nwtail}). 

\subsubsection{Spectral Modeling}
\label{sec:specmod}

We extracted spectra from regions based on the 
temperature map features for detailed fitting using XSpec 11.3.0. 
These regions are overlaid on the temperature and surface
brightness maps in the upper right and 
lower right panels of Figure \ref{fig:Fepeaktmap}, respectively, 
and are listed in Table \ref{tab:n6868tmapspecreg} for NGC\,6868 and 
Table \ref{tab:n6861specreg} for NGC\,6861. The data were binned to 
ensure a minimum of $20$ counts per bin and also using constant logarithmic 
bin size. We found no significance differences in the results obtained 
from these two binning methods. Backgrounds were taken from  
rescaled blank sky background sets (see  \S\ref{sec:obs}), extracted from 
the same region as each source.  We model the spectra  using absorbed 
APEC thermal plasma (Smith \etal 2001 ) or APEC plus power law models 
and fit the data over the $0.5-7$\,keV energy range. To check our 
sensitivity to the soft Galactic background, we refit the data over the 
$0.7-7$\,keV range where this soft Galactic background will not 
contribute and found the fits unchanged. 

\noindent{\bf NGC\,6868:} 

For NGC\,6868 we concentrate on
features in and to the south of the galaxy. We extract spectra for
the nuclear  region of the galaxy (N), the lowest temperature region
east of NGC\,6868's  center (cold clump C), the filament of higher
temperature gas to the southwest (F) and the lower temperature, spiral 
 tail-like feature (the combined regions T). The results of our spectral 
fits for NGC\,6868, listed in Table \ref{tab:n6868tmapspec}, generally 
confirm the results of Figure \ref{fig:Fepeaktmap}. The galaxy nucleus
is well described by a two component APEC + power law model with 
gas temperature $0.63^{+0.07}_{-0.06}$\,keV, solar abundance, and 
power law index $\sim 1.4$ to account for unresolved X-ray binaries in
the central region of the elliptical galaxy and/or a weak AGN. 
Outside NGC\,6868's
central region, the spectrum of the higher temperature  
filament (region F) and the southern spiral-like  tail (region T) 
are both well described by single APEC thermal plasma models with  
temperatures of $0.70 \pm 0.05$ and $0.62 \pm 0.04$\,keV,
respectively. APEC models for the
filament F have a modestly improved reduced $\chi^2$ for low 
abundances ($\sim 0.3-0.2\Zs$), suggesting that the filament 
may be composed of IGM rather than galaxy gas. 
The most problematical region to interpret in the temperature 
map of NGC\,6868 is the cool clump region C. While a statistically 
acceptable fit (${\rm null = 0.09}$) is found for a single temperature 
($kT= 0.59^{+0.04}_{-0.05}$\,keV) APEC model, there are significant
residuals in the $0.7-1$\,keV region and 
the $\chi^2/{\rm dof} = 38/28$  is not particularly good. 
Addition of a second APEC or power law component did not improve the fit. 

\noindent{\bf NGC\,6861:}

To determine the thermodynamics properties of gas surrounding 
the spheroidal galaxy NGC\,6861, we measure the gas temperature 
in three  rectangular regions: region E located $43.2$\,kpc to
the east of NGC\,6861's nucleus in the direction of NGC\,6868, 
region NW located to the northwest coincident with the 
northwestern tail shown in Figure
\ref{fig:nwtail} and in  the lower right panel of 
Fig. \ref{fig:Fepeaktmap}, and region WT containing the bright 
tail to the west of NGC\,6861. We also model spectra within 
NGC\,6861 using an elliptical region G
centered on NGC\,6861's nucleus with semi-major (minor) axes of $4.2$
($2.6$)\,kpc, respectively, containing the
nucleus and brightest galactic emission (see
Fig. \ref{fig:n6861sbfits}), and in a forward annular sector
FA, with inner and outer radii of $5.3$\,kpc 
and $14.3$\,kpc, respectively, and constrained to lie between $80^\circ$ and
$282^\circ$ measured counter-clockwise from west, containing the 
plume (Fig. \ref{fig:nwtail}) and 
 `sheath', identified in Figure \ref{fig:Fepeaktmap}, surrounding 
the galaxy's central region. The results of our spectral fits to these 
regions are listed in Table \ref{tab:n6861specfits}. The central
region of the galaxy (G) is well modeled by
a $0.66^{+0.11}_{-0.07}$\,keV APEC model, describing the contribution
of  the diffuse galaxy gas, plus a power law component with photon 
index $1.86 \pm 0.6$, consistent with an active nucleus or 
population of unresolved X-ray binaries. The temperature of gas in 
the forward annulus region (FA) 
is significantly higher ($1.27 \pm 0.15$\,keV), in agreement with
Figure \ref{fig:Fepeaktmap}, and higher  metal
abundances ($\gtrsim 0.5\Zs$) are favored, as expected for 
gas associated with the galaxy. Gas in the western tail is well 
described by a single temperature APEC model, with gas temperature 
$kT = 1.18^{+0.10}_{-0.12}$ and abundance $0.25^{+0.21}_{-0.13}\Zs$, 
confirming that the temperature in the western tail (WT) is also
high. The low best fit metallicity in region WT, albeit with large 
uncertainties, favors a large IGM component rather than a purely 
galaxy origin for the gas in the western tail.
 
The properties of the X-ray emission to the northwest and east of NGC\,6861 
are more uncertain. The temperature and abundance of gas to the
northwest of NGC\,6861, along the 
possible northwestern tail, are $0.90^{+0.09}_{-0.1}$\,keV and
$0.22^{+0.26}_{-0.11}\Zs$, with similar abundance but lower
temperature than gas in the western tail. 
In contrast, a single temperature APEC model
fit to the spectrum in region E east of NGC\,6861 
(denoted fit E1 in Table \ref{tab:n6861specfits}) gives
a low gas temperature ($\sim 0.62 ^{+0.14}_{-0.11}$\,keV), 
consistent at the $90\%$ CL
with the gas temperature to the north of NGC\,6868. However, the 
best fit abundance is anomalously low ($\lesssim 0.06\Zs$) in that
model when compared to metallicities outside dominant galaxies in 
other galaxy groups,  and the residuals show a  
'see-saw' pattern often associated with multi-temperature gas 
(Buote 2000). A two-component APEC model (denoted E2 in Table 
\ref{tab:n6861specfits}) with a more realistic 
abundance value of $0.2\Zs$, consistent with the best fit abundances
found in the northwest and western tail regions, yields  gas temperatures of
$\sim 0.46^{+0.17}_{-0.13}$\,keV and $\sim 1.24^{+0.31}_{-0.17}$\,keV,
but the  reduced $\chi^2$ is unchanged over that for the single APEC
fit. Deeper observations are required to disentangle the complex
temperature structure to the east of NGC\,6861.
   
\subsubsection{NGC\,6861's Tails: Stripping or Sloshing?}

Since bifurcated tails are common features in 
ram-pressure stripped galaxies, such as those  infalling
into the Virgo cluster (see, e.g. Randall \etal 2008), 
NGC\,6861's bright western and fainter northwestern 
tails might suggest that it is infalling towards NGC\,6868 
near the core of the AS0851 galaxy group.  However, most 
ram-pressure stripped tails are composed of galaxy gas with higher
metallicities and lower gas temperatures than the surrounding group
IGM, in contradiction with the $\gtrsim 1$\,keV gas temperatures and 
low ($\sim 0.2$) best-fit abundances  measured  in NGC\,6861's tails. 
Long, hot tails can be formed by turbulent-viscous stripping of 
galaxy gas and its subsequent mixing and thermalization with the group IGM 
during supersonic motion of the galaxy through the ambient group
gas, or by gravitational focusing of the IGM into an adiabatically heated 
 Bondi-Hoyle wake (see, e.g. Machacek \etal 2005b). 
However, when such supersonic infall occurs in the plane of the sky,
as we expect here (see \S\ref{sec:merger}), we should see a rise in 
temperature and density along a narrow shock front preceding the
motion, followed by a cold front edge closer to the galaxy. We do see 
higher temperature gas 
($kT \sim 1.3^{+0.3}_{-0.2}$\,keV), compared to the ambient IGM 
($\sim 0.7-0.9$\,keV), in the sheath surrounding NGC\,6861 
(region FA in Fig. \ref{fig:Fepeaktmap}). If the factor $\sim 1.9$ 
temperature rise in the sheath is due to a shock, we would expect to
see a corresponding factor $\sim 2$ increase in density 
from the Rankine-Hugoniot conditions and, thus, a surface brightness 
increase by a factor $4$ in the same region (see, e.g. Landau \& Lifshitz
1959). Although we find no evidence in these data
for the sharp change in surface brightness (and thus density) that
would  be associated with such a shock, nor do we see the associated 
merger cold front expected if NGC\,6861 were moving 
supersonically in the plane of the sky through the ambient IGM, 
projection effects in these sparse data may make such features 
difficult to observe.

A more promising scenario may be that surface brightness and temperature
asymmetries are the result of a gravitational encounter between 
NGC\,6868 and NGC\,6861, each the dominant galaxy in their own
subgroup.  Then the dark matter distribution is bimodal,  
with each galaxy residing near its own local minimum of the 
gravitational potential. This would explain why
double $\beta$-models provide a good description of the mean surface
brightness profile around each galaxy. However, since the merger is
likely occuring in the plane of the sky, this subgroup structure may
be difficult to discern from the optical galaxy distributions. 
We may be viewing the encounter after 
the subgroups have passed through pericenter, but before the gas halos
have been completely stripped and merged.   The  strong temperature 
and surface brightness asymmetries around each of the merging partners
may be due to non-hydostatic motions of the gas (`sloshing'), 
as gas is displaced from the center of each subcluster's dark matter 
potential, similar to those 
produced in simulations of premerger encounters of more massive
subcluster halos. When gas is present in both halos, cold
fronts with cool spiral-shaped tails may form about one subcluster,
while higher temperature shock-heated features may form around the
other (see  Figure 9 of Ascasibar \& Markevitch 2006). These
simulations suggest that the asymmetries persist for gigayears 
after the encounter. Thus the observation of
sloshing  signatures around each galaxy, even though they are separated
by $207$\,kpc (sound crossing time of $500$\,Myr), is reasonable.

\subsection{The Case for AGN Feedback}
\label{sec:cavity}

AGN activity and black hole growth may be influenced by either ram pressure, 
through backflow of partially stripped material  onto the 
central regions of the galaxy,  or during a close gravitational
encounter between two galaxies, that would also produce sloshing. 
Since both NGC\,6868 and NGC\,6861 currently host weak radio sources
at their nuclei consistent with AGN, we look for signatures in the
X-ray gas that might signal past episodes of more energetic AGN activity.  

\begin{figure}[t]
\begin{center}
\includegraphics[height=2.56in,width=3.0in,angle=0]{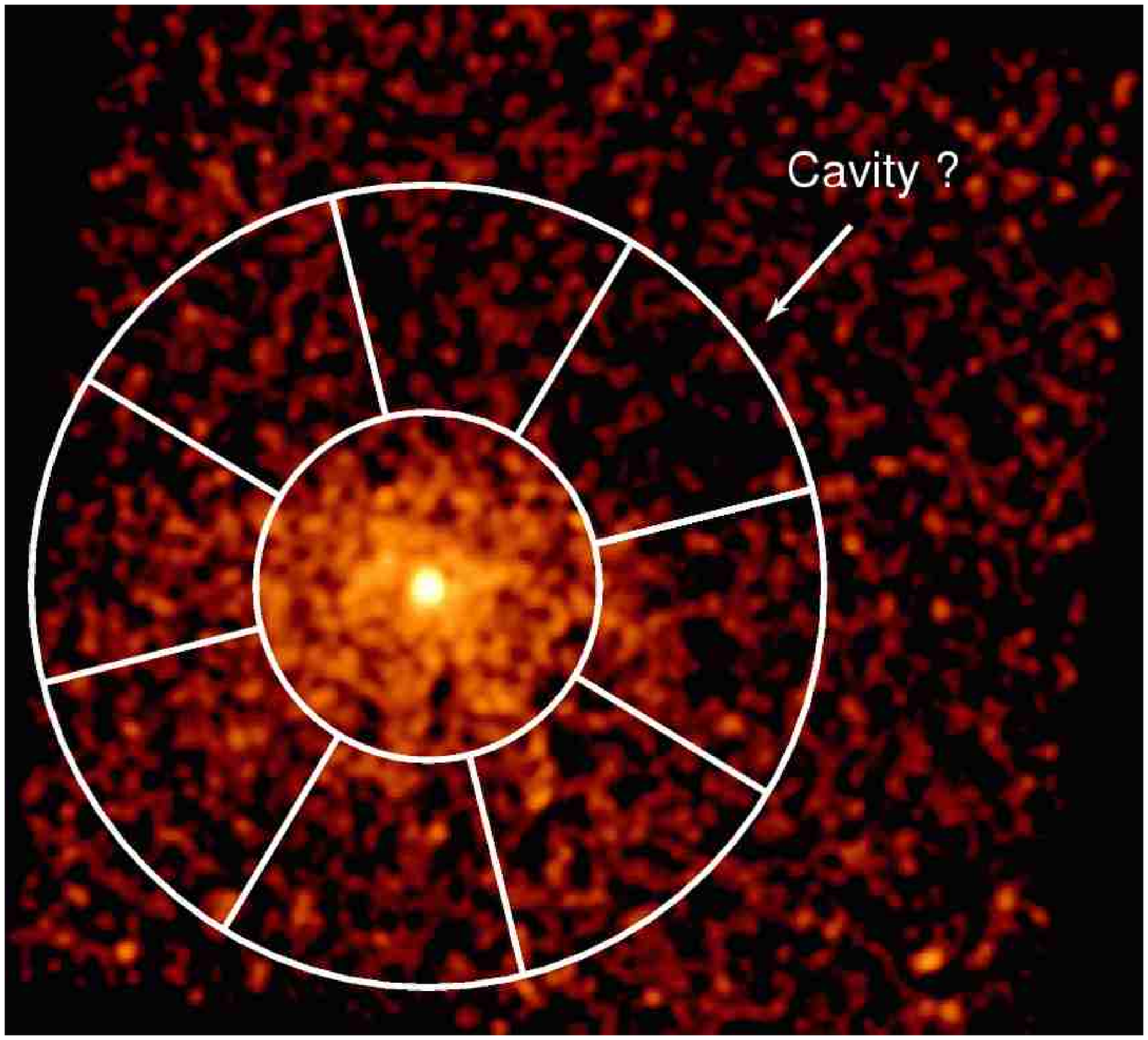}
\includegraphics[height=3.0in,width=2.37in,angle=270]{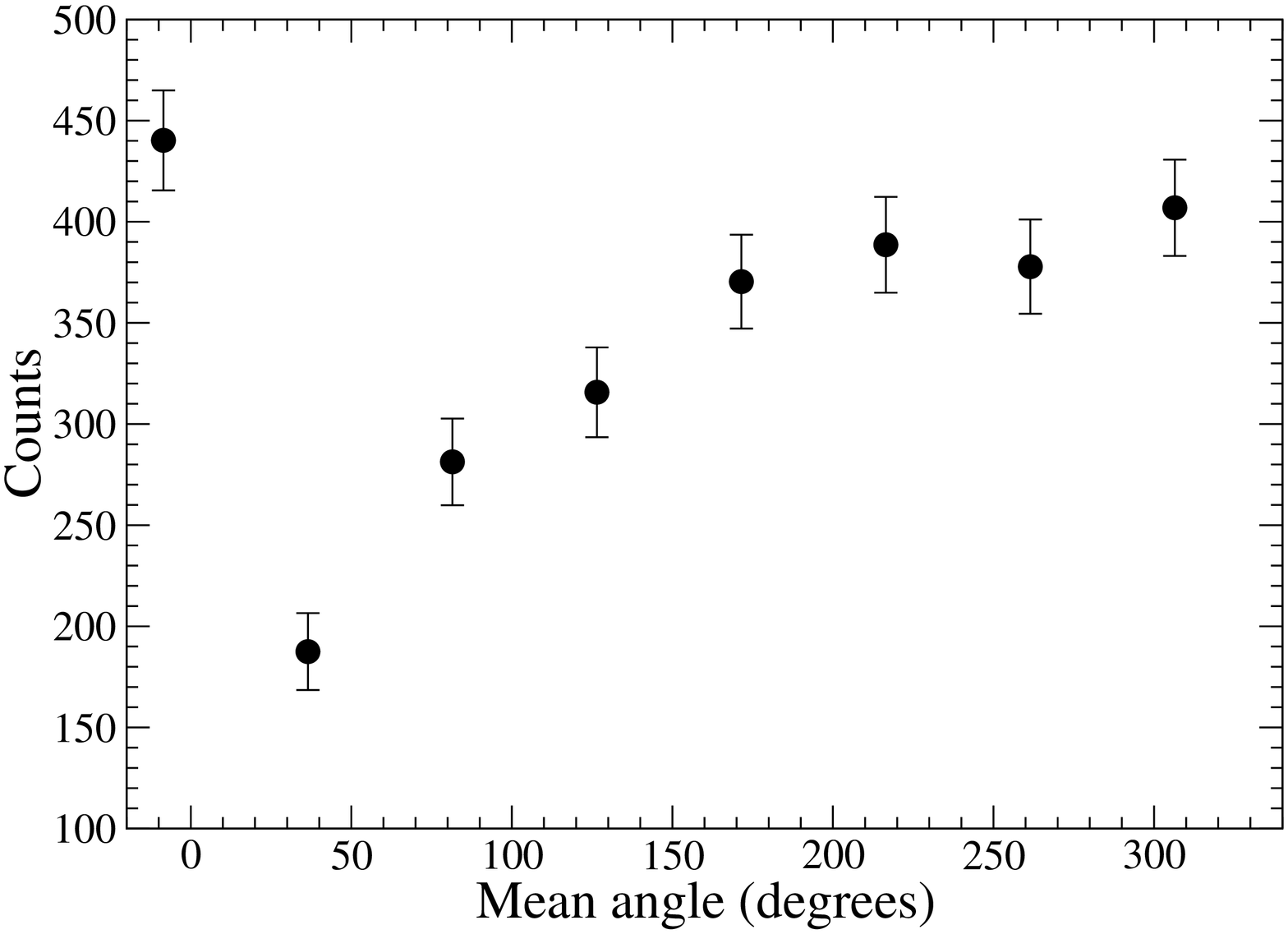}
\caption{\footnotesize{({\it upper}) $0.5-2$\,keV blank-sky 
 background subtracted, exposure corrected {\it  Chandra} image of 
  NGC\,6868 with a panda region overlaid to identify X-ray surface 
  brightness deficits (cavities) in the group IGM. 
$1 \,{\rm pixel} = 1\farcs968 \times 1\farcs968$ and the image 
 has been smoothed with a $6"$ Gaussian kernel. The annular sectors
 are  centered on NGC\,6868's nucleus ($20^h09^m54.1^s$,$-48^\circ22'46.4"$)
 with inner and outer radii of $168''$ ($22.5$\,kpc) and $387\farcs6$ 
 ($51.9$\,kpc), respectively. North is up and east is to the left. 
({\it lower}) Net source counts as a function of the mean angle 
$\phi$ of the annular sector, measured counterclockwise from west. 
Error bars represent  $1\sigma$ uncertainties in the source counts. 
Note the surface brightness deficit at $\phi = 36.5^\circ$.}}
\label{fig:n6868cavity}
\end{center}
\end{figure}

\subsubsection{`Ghost' Cavities Near NGC\,6868?}
\label{sec:n6868cavites}

Figures \ref{fig:chandramosaic} and  \ref{fig:n6868cavity} 
show a roughly circular deficit in X-ray surface brightness  
centered $4\farcm9$ (in projection) to the northwest of
NGC\,6868's nucleus. In the upper panel of 
Figure \ref{fig:n6868cavity} we use $8$ annular
sectors, centered on the nucleus of NGC\,6868
and with inner and outer radii of $168''$ ($22.5$\,kpc) and $387\farcs6$
($51.9$\,kpc), respectively, to investigate more quantitatively 
this northwest cavity. We denote the position of each sector by the
mean angle $\phi$ for the sector measured counterclockwise from the 
$+x$ axis (west). In the lower panel we plot the number of
 net source counts in each annular sector in the $0.5-2$\,keV energy band 
as a function of sector position.
Error bars represent $1\sigma$ uncertainties in the net source
counts. We find that the number of net counts ($187 \pm 19$) in the 
sector coincident with the northwest cavity 
($\phi =37.5^\circ$) is lower by  $\gtrsim 3\sigma$ than the counts in
its neighboring sectors, ($281.3 \pm 21.4$ counts at $\phi =
81.5^\circ$ and $440.2 \pm 24.7$ counts at $\phi=-8.5^\circ
[351.5^\circ]$). 

The cavity candidate is a `ghost' cavity, since no radio emission has 
as yet been observed in this region, has low contrast and no well defined rim, 
similar to the radio-faint cavity in 2A 0335+096 (Birzan \etal 2004). 
Features of this kind need to be confirmed in deeper 
X-ray observations and explored with lower frequency radio
observations to detect the radio plasma expected to reside in their
interiors.  Furthermore, the projected spherical morphology is
somewhat puzzling given the evidence for gas motions in
\S\ref{sec:n6868edge} and \S\ref{sec:temp} that might 
distort the cavity's shape. However, assuming the cavity is an 
evolved  buoyant bubble filled with relativistic radio plasma from 
past AGN activity, we may use the geometry of the cavity and the 
mean pressure in the surrounding IGM to estimate the enthalpy 
($4pV$) carried in the bubble. We calculate the mean pressure 
of gas that was evacuated to
form the bubble from the density model in eq. \ref{eq:densemod} and the
temperature of the IGM given in Table \ref{tab:n6868edgespec}.   
Assuming a spherical bubble of radius $11.7$\,kpc 
($87''$) oriented close to the plane of the sky, we find the enthalpy
carried by the bubble of $4pV \sim 5.9 \times 10^{56}$\,ergs, similar
to ghost cavities in Abell\,262 (Blanton \etal 2004) or 
HCG\,62 (Birzan \etal 2004) and to
younger cavities in galaxies, such as M87 (Young \etal 2002; 
Forman \etal 2005) and NGC\,4472 (Biller \etal
2004). 
 
The age of the bubble may be estimated by assuming that (1) the bubble
rises with the speed of sound $c_s$ in the ambient medium $t = d/c_s$, 
(2) the bubble rises buoyantly at the terminal velocity $v_t$ such that 
$t = d/v_t \sim d \sqrt{SC/2gV}$, where $V$ is the volume of the
bubble, $S$ its cross-section, and $C=0.75$ the drag coefficient, or  
(3) the time $t=2\sqrt{r/g}$ for gas to refill the displaced volume as
the bubble moves upward, where $r$ is the radius of the cavity and $g$ the 
acceleration of gravity (Churazov \etal 2001; McNamara \etal 2000;
Nulsen  \etal 2002; Birzan \etal 2004). For the northwestern bubble
these imply bubble ages of $88$, $107$, $119$\,Myr, respectively, for
the three methods, and a mechanical luminosity of $L_{\rm mech} \sim 1.8
\times 10^{41}$\ergs.  Bubble ages of $\sim 10^8$\,yr are  
typical for ghost bubbles in other systems (see, e.g. Birzan \etal 2004). 

A partially rimmed, spherical X-ray surface brightness feature, with 
radius $\sim 60"$ may also be seen in Figure \ref{fig:chandramosaic} 
located $3\farcm1$ to the southwest of NGC\,6868's nucleus. A filament
of hot gas (region F in Fig. \ref{fig:Fepeaktmap}) is seen to cross
the candidate cavity. If this feature is also a buoyant ghost bubble, 
perhaps broken by the hydrodynamic motions in the gas caused by 
interactions between NGC\,6868 and NGC\,6861, we estimate its 
enthalpy, age and mechanical power to be $4 \times 10^{56}$\,ergs, 
$55 - 78$\,Myr, and $\sim 2 \times 10^{41}$\ergs, respectively, 
comparable to that of the northwest cavity.

\subsubsection{AGN Heating in NGC\,6861?}

The surface brightness in the plume rising to the northeast from NGC\,6861 
and in the western tail increase by factors $\sim 2 $ and 
$\sim 3$, compared to regions to the southeast at comparable distance
from NGC\,6861's nucleus, and have a  $\sim 0.5$\,keV higher
temperature than the central region of that galaxy 
(see Fig. \ref{fig:Fepeaktmap} and regions FA and WT
compared to G in Table \ref{tab:n6861specfits}). One possible explanation
is that gas in these regions may have been disturbed and heated by 
AGN activity. Assuming that the western tail is contained in a 
cylindrical volume oriented in the plane of the sky with 
radius $18$\,kpc and length $37.7$\,kpc (based on 
the WT spectral region), we find the tail contains 
$\sim 2.6 \times 10^{9}\eta^{1/2}\,\Ms$ of hot gas. Since the gas in the 
tail appears clumpy, the filling factor $\eta$ is likely less than one, 
so that this is an upper bound on the actual mass of hot gas in the
tail. The energy necessary to raise the temperature of the gas in the 
tail from $\sim 0.7$\,keV found for gas in the bright central region of the
galaxy (region G) or the surrounding IGM to $1.18$\,kev measured in 
the tail is $\sim 6 \times 10^{57}$\,ergs. 
Outburst energies of up to $10^{61}$\,ergs 
have been measured from shocks observed in, for example, Hercules A 
(Nulsen \etal 2005a) and Hydra A (Nulsen \etal 2005b), while outburst 
energies $\sim 10^{58}$\,ergs have been measured in M87 (Forman \etal
2005), whose $3 \times 10^9\,\Ms$ central black hole has 
mass comparable to that inferred for NGC\,6861 from the 
$M_{\rm BH}-\sigma_*$ relation. 
Thus energetic outbursts from a  $2.5 \times 10^9\,\Ms$ black hole at 
the center of NGC\,6861 would  have sufficient energy to heat 
and produce turbulent motions in the gas, that might also contribute to the  
disturbed gas morphology observed near NGC\,6861. 

\section{On NGC\,6861's Black Hole Mass}
\label{sec:discuss}

The discrepancy between the black hole mass prediction from the  
Magorrian relation (eq. \ref{eq:magorrian}) and the 
$M_{\rm BH}-\sigma_*$ relation (eq. \ref{eq:mbhsigma}) is still 
difficult to understand. If NGC\,6861's black hole mass is large, then
either an existing massive stellar bulge has been stripped from
NGC\,6861 or the coevolution of NGC\,6861's stellar bulge has been
inhibited or delayed during a period of rapid black hole growth.  
In the first case, hydrodynamic processes such as ram pressure
and turbulent viscous stripping, plasma outflows from AGN, or gas 
sloshing may heat and redistribute the gas in and around NGC\,6861,
but would be  ineffective at changing the galaxy's stellar
distribution. Stars in the outer halo could be stripped tidally if the 
impact parameter of the collision between NGC\,6861 with NGC\,6868
is sufficiently small, as observed in dumbbell galaxy systems 
such as NGC\,4782/NGC\,4783.  However, even in the deeply penetrating
collisions of such dumbbell galaxies, 
simulations show that only $10-20\%$ of the stellar mass is expected
to be stripped (Madejsky \& Bien 1993).  
For a $2.5 \times 10^9\,\Ms$ black hole, equation
\ref{eq:magorrian} predicts a stellar bulge mass for NGC\,6861 of 
$\sim 10^{12}\,\Ms$, such that previous interactions would
have had to strip $\sim 90\%$ of the stellar mass to be in agreement
with the K-band mass estimates given in Table \ref{tab:bhmass}. 
One would expect to see evidence for such cataclysmic tidal
disruption in tidal bridges, tails or other stellar morphological 
distortions, that are not observed in the stellar
distributions of either NGC\,6861 or NGC\,6868. Also this scenario
would imply the direct passage of the subgroup gas cores through each 
other and much more disruption in the X-ray gas in NGC\,6868 than 
is observed.

The second case, in which the growth of the stellar bulge is
inhibited, may be more likely. NGC\,6861 shows evidence for a stellar
bar (Koprolin \& Zeilinger 2000). Stellar bars are likely to be 
formed in mergers and allow angular momentum of infalling matter 
to be dissipated. Gas and dust may then be transported along the
bar to the galaxy's nuclear region to fuel one or several episodes 
of rapid black hole growth. If this accretion event triggers an 
AGN outburst in NGC\,6861, gas otherwise available for star 
formation could be heated and inflated, delaying the expected growth 
of the bulge (Silk \& Rees 1998). The total energy deposited into the 
ambient gas from an episode of rapid black hole growth  is 
$f\epsilon\Delta Mc^2$, where $\Delta Mc^2$ is the rest mass energy of
the accreted matter, $\epsilon$ is the fraction of the rest mass
energy released in the accretion event, and $f$, the feedback efficiency, 
measures how efficiently that energy is coupled to the surrounding
galaxy and subgroup gas. For the standard assumption of radiative 
efficient accretion ($\epsilon = 0.1$) and efficient feedback 
$f \gtrsim 0.05$ (Di Matteo \etal 2008; Churazov \etal 2002), the
total energy deposited in the surrounding gas as the black hole grows 
from $2.1 \times 10^8\Ms$ predicted by the Margorrian relation to 
$2.5 \times 10^9\Ms$, consistent with the $M_{\rm BH}-\sigma_*$ relation,
would be large ($\gtrsim 2 \times 10^{61}$\,ergs). This is comparable to the 
most energetic AGN outbursts observed in massive clusters (see, e.g. Nulsen
\etal 2005a for Hercules A; McNamara \etal 2002 for MS 0735.6+7421;
Nulsen \etal 2005b for Hydra A). 
If accretion proceeds at the Eddington rate, 
\begin{equation} 
 \dot{M_{\rm Edd}} = (0.22/\epsilon)(M_{\rm BH}/10^8)\,\Ms\,{\rm
   yr}^{-1}\,\,\,,
\label{eq:accrete}
\end{equation}
the accretion timescale would be $\sim 110$\,Myr and
the mean radiative power, $\sim 10^{47}$\ergs. The power that couples to the 
surrounding gas ($\sim 5 \times 10^{45}$\ergs) is more than three orders of
magnitude larger than that needed to balance X-ray cooling in the 
galaxy. Such an event would likely completely disrupt the gas in
NGC\,6861 and its subgroup. However the accretion and feedback
efficiencies are highly uncertain. If the growth 
of NGC\,6861's black hole was triggered by the ongoing merger with the
NGC\,6868 subgroup in AS0851, either the accretion is radiatively inefficient 
($\epsilon << 0.1$), the released rest-mass energy is only weakly 
coupled to the galaxy and subgroup gas ($f << 0.05$; see, e.g.,  
Kurosawa \etal 2009), and/or the 
black hole grows through multiple, less energetic outburst events over
a much longer timescale.  
For example, if $f\epsilon \sim 10^{-6}$, the energy from AGN activity
deposited in the  surrounding subgroup gas from this episode of rapid 
black hole growth would be $\sim 10^{58}$\,ergs, sufficient to heat 
the western tail. Alternatively, 
the period of rapid black hole growth may have occurred at very early
times, before NGC\,6861 was assembled into the current 
galaxy group/subgroup structures in AS0851. Then a very strong AGN
outburst, as indicated above, may have swept gas from the galaxy, 
delaying bulge formation by quenching star formation and heating or
blowing away any surrounding subgroup gas. 
Growth of the associated stellar bulge for NGC\,6861 may only now be 
occurring, not through star formation, but by the cannibalizaton of the 
stellar distributions already in place as NGC\,6861 merges with nearby
galaxies in the group and ultimately NGC\,6868 (Silk \& Rees 1998).

A third possibility is that the Magorrian relation holds and the
central black hole in NGC\,6861 is $\sim 2 \times 10^8\,\Ms$, typical
for early type galaxies in groups. 
Then gravitational interactions between NGC\,6861 and NGC\,6868 or
other group galaxies 
would also need to increase the central 
stellar velocity dispersion in NGC\,6861 to anomalously high values. 
Strong tidal forces in a deeply penetrating encounter  
may distort stellar velocity 
profiles, causing them to increase with distance from the galaxy
center (see, e.g. Madejsky \& Bien 1993 for NGC\,4782). For finite
apertures, this might contaminate the measurement of
the central stellar velocity dispersion and thus cause the black hole 
mass to be overestimated. For example, the measurement of $382 \pm
7$\kms by Koprolin \& Zeilinger (2000) is the average stellar 
velocity dispersion within $5''$ ($670$\,pc) of the galactic center,
well outside the sphere of influence of the supermassive black hole. 
However, if the velocity dispersion profile has been affected by tidal
forces in a merger, as in NGC\,4782, one would expect that the stellar 
velocity dispersion would increase when averaged over larger radii, 
which is not observed in NGC\,6861 (Koprolin \& Zeilinger 2000). 
It is also difficult to understand in this scenario why the stellar
velocity dispersion of NGC\,6861 would be so profoundly affected by
the interaction, while that of its likely interacting partner 
NGC\,6868 is not. If the Magorrian relation holds, the anomalously 
high stellar velocity dispersion in 
NGC\,6861 may require NGC\,6861 to have experienced one or more recent 
minor mergers unrelated to the interaction with NGC\,6868 that is 
responsible for the large scale gas sloshing observed in the group.  

\section{Conclusions}
\label{sec:conclude}

In this paper we used {\it Chandra} observations of NGC\,6868 and 
NGC\,6861 in the AS0851 galaxy group to measure the density and
temperature structure of gas in these two dominant group elliptical
galaxies. This system is particularly important because, 
although the brightest
group galaxy has a black hole mass $M_{\rm BH} \sim 3.3 \times
10^8\,\Ms$ predicted by the $M_{\rm BH}-\sigma_*$ relation,  that is 
both consistent with the Margorrian $M_{\rm BH}-M_{\rm bulge}$
relation and typical of black hole masses inferred for the dominant 
galaxies in other cool groups, the central black hole mass   
inferred from the $M_{\rm BH}-\sigma_*$ relation for the second
brightest group galaxy NGC\,6861 ($M_{\rm BH} \sim 2.5 \times
10^9\,\Ms$) is more than an order of magnitude higher than that 
predicted by the Magorrian relation. The goal of this work was to use 
these X-ray data to explore possible interaction scenarios 
(infall and gas stripping, gas sloshing, AGN outburst activity) 
that might affect galaxy evolution and black hole growth. 
 
We find: 
\begin {itemize}

\item {The X-ray surface brightness distributions around NGC\,6868 and
  NGC\,6861 are each well described by double $\beta$-models with
  inner (outer) beta and core radii of $0.65$ and $0.49$\,kpc ($0.52$
  and $11$\,kpc) for NGC\,6868 and $0.61$ and $0.5$\,kpc ($0.38$
  and $11.9$\,kpc) for NGC\,6861. This suggests that NGC\,6868 and
  NGC\,6861 may each be the dominant galaxy in a galaxy subgroup that
  is merging. Lack of evidence for  substructure in the galaxy
  reshift distribution suggests that we would be viewing such a
  subgroup merger close to the plane of the sky.  
}

\item{ X-ray surface brightness and temperature maps of NGC\,6868 show
    a cold front $23 \pm 3$\,kpc to the north of the galaxy nucleus
    and a spiral-shaped tail of cool $0.62$\,keV gas to the south,
    suggesting relative motion between the galaxy and the surrounding
    IGM. Analysis of the density and temperature across the edge 
    constrains the relative motion between NGC\,6868 and the IGM to be 
    at most transonic ($\lesssim 460$\kms).  While we cannot rule out
    the possibility that the tail is composed of ram-pressure stripped
    galaxy gas, the tightly wound spiral-like morphology of the tail 
    in the temperature map strongly suggests that that the tail is a 
    result of the `sloshing' of gas in the gravitational potential, 
    set in motion by the previous gravitational encounter between the 
    NGC\,6868 and NGC\,6861 subgroups.
}

\item{X-ray surface brightness and temperature maps for NGC\,6861, 
$206$\,kpc to the west of NGC\,6868, show a complex morphology 
also suggestive of a subgroup merger.  While spectra of the central 
$\sim 5$\,kpc of NGC\,6861 containing the bright stellar disk is 
well described by a cool $0.66^{+0.11}_{-0.07}$\,keV thermal plasma model
plus a power law with photon index $-1.87$ for the central active
galactic nucleus (AGN), the temperature of gas in the $5-14$\,kpc `sheath' 
outside NGC\,6861's central region is a factor $\sim 2$ higher. A 
hot ($0.9 - 1.2$\,keV), bifurcated tail is seen trailing NGC\,6861 $38$\,kpc 
to the west and northwest. These features may suggest that 
the NGC\,6861 subgroup is undergoing ram-pressure or 
turbulent-viscous stripping during supersonic infall towards
NGC\,6868. However, comparing the density and temperature morphologies 
of NGC\,6868 and NGC\,6861 taken together with simulations from the 
literature, the more likely explanation is that the hot features in 
NGC\,6861 are the consequence of gas sloshing caused by the ongoing 
merger of two subgroups, whose respective hot gas halos have not yet 
been stripped. 
}

\item{NGC\,6868 may have undergone a period of merger-induced AGN
    activity. X-ray surface brightness images show weak evidence for
roughly circular surface brightness deficits to the northwest and
southwest of NGC\,6868. If confirmed in deeper X-ray images or low
frequency radio observations to be buoyant bubbles  
filled with relativistic plasma from previous AGN activity, the bubbles' 
enthalpies ($5 \times 10^{56}$\,ergs), mechanical power ($2 \times
10^{41}$\ergs) and ages ($\sim 60 - 100$\,Myr) are typical for `ghost' 
cavities observed in other systems.} 

\end{itemize}

 While the X-ray data favor a sloshing interaction scenario  
aided by AGN activity to explain possible cavities near NGC\,6868 and 
gas heating in/near NGC\,6861, an explanation for the discrepancy
between the Magorrian and $M_{\rm BH}-\sigma_*$ relation predictions 
for the black hole mass at the center of NGC\,6861 remains challenging. 
Deeper X-ray observations that probe closer to the central engine and 
map with greater precision the complex density and temperature
structure of the gas are needed to constrain the gasdynamics and 
orbital parameters of the merger and determine the energies and 
timescales of associated AGN activity. More detailed theoretical
modeling in gravity + hydrodynamic numerical simulations that are guided by
those data may then be able to determine
how mergers affect the high mass end of the black hole mass relations
and solve the mystery of NGC\,6861's black hole mass. 

\acknowledgements

This work is supported in part by NASA grant NNX07AH65G  
and the Smithsonian Institution. 
This work has made use of the NASA/IPAC Extragalactic Database (NED)
which is operated by the Jet Propulsion Laboratory, California
Institute of Technology,  under contract with the National
Aeronautics and Space Administration. We wish to thank John Huchra 
for help using the Harvard-Smithsonian CfA Redshift Survey (ZCAT) 
database for galaxy redshift data, Ryan Johnson and 
Paul Nulsen for helpful discussions, and an anonymous referee for
helpful comments.

\noindent{\it Facilities:} CXO (ACIS-I)

\begin{small}

\end{small}

\begin{deluxetable}{cccccc}
\tablewidth{0pc}
\tablecaption{Galaxy Properties and Black Hole Masses\label{tab:bhmass}}
\tablehead{\colhead{Name} & \colhead{$K_s$(total)} & \colhead{$M_\star$} & 
\colhead{$\sigma_\star$} &\colhead{$M^{\sigma}_{\rm BH}$} & 
\colhead{$M^\star_{\rm BH}$} \\  
 &  & ($10^{11}\Ms$)  & (km\,s$^{-1}$)  & ($10^8\Ms$) & ($10^8\Ms$) } 
\startdata
NGC\,6868 & $7.317$ & $1.83$ & $250 \pm 10$ & $3.3$ & $3.1$ \\
NGC\,6861 & $7.708$ & $1.28$ & $414 \pm 17$ & $25$ &  $2.1$ \\
\enddata
\tablecomments{\footnotesize{ Columns (1) galaxy name; (2) total 2MASS K-band
  luminosity (NED); (3) stellar mass (bulge mass) estimated 
  using K-band luminosities (Gilfanov 2004); (4) stellar velocity 
  dispersion (Wegner \etal 2003); (5) and (6) 
  central black hole mass from the $M_{\rm BH} -\sigma_*$ 
  (eq. \protect\ref{eq:mbhsigma}; Tremaine \etal 2002) and 
  the $M_{\rm BH}-M_{\rm bulge}$ 
  (eq. \protect\ref{eq:magorrian}; H\"{a}ring \& Rix 2004) relations, 
  respectively. }
}
\end{deluxetable}

\begin{deluxetable}{cccccc}
\tablewidth{0pc}
\tablecaption{Density Model Fits to NGC\,6868's Edge\label{tab:n6868edgefits}}
\tablehead{\colhead{Fit range} & \colhead{$\alpha_1$} &
  \colhead{$r_{\rm edge}$} & \colhead{$J$} & \colhead{$n_1$} &\colhead{$n_2$} \\
 (kpc) & & (kpc) & & ($10^{-3}$\cmc) & ($10^{-3}$\cmc)}
\startdata
$[7.9,34.0]$ & $-0.40$ & $22.7$ & $2.2$ &$1.52$ &$0.84$ \\
$[7.9,68.0]$ & $-0.15$ & $20.3$ & $1.8$ &$1.47$ &$1.00$ \\
$[5.8,34.0]$ & $-0.82$ & $26.5$ & $2.6$ &$1.38$ &$0.66$ \\
$[5.8,68.0]$ & $-0.64$ & $22.5$ & $1.7$ &$1.16$ &$0.85$ \\
\enddata
\tablecomments{\footnotesize{Density models fits across the $0.5-2$\,keV
    surface brightness edge in NGC\,6868 shown in
    Figs. \protect\ref{fig:chandramosaic} and 
   \protect\ref{fig:n6868edgefit} for the density model given in 
     eq. \protect\ref{eq:densemod}. $n_1$ and $n_2$ are electron
     densities inside and outside the edge, respectively, assuming
     ($kT, A$) of ($0.64$\,keV, $0.5\Zs$) and ($0.72$\,keV, $0.3\Zs$) 
    inside and to the north outside the edge, respectively. 
See Table \protect\ref{tab:n6868edgespec}.} 
}
\end{deluxetable}

\begin{deluxetable}{ccccc}
\tablewidth{0pc}
\tablecaption{NGC\,6868 Spectral Regions Across the Edge\label{tab:n6868edgespecreg}}
\tablehead{\colhead{label} & \colhead{shape} & \colhead{center} &
  \colhead{radius$^a$}&\colhead{angle} \\
& & (RA, Dec) & (arcsec) & (deg) 
}
\startdata
NEdge$_{-1}$ &elliptical sector &$20:09:54.1,-48:22:46.4$ & $82.4,131.9$ & $6$  \\
NEdge$_{+0}$ & elliptical sector &$20:09:54.1,-48:22:46.4$ &$131.9,211.$ &$6$ \\
NEdge$_{+1}$ & elliptical sector &$20:09:54.1,-48:22:46.4$ &$211.,337.6$ & $6$ \\
S$_{+1}$ & elliptical sector &$20:09:54.1,-48:22:46.4$ &$211.,337.6$ &$6$\\
\enddata
\tablecomments{\footnotesize{NEdge regions are concentric to the
    bounding ellipse and lie within the sector 
shown in  Fig. \protect\ref{fig:edgereg} and described in 
\protect\S\ref{sec:n6868edge}. S$_{+1}$ uses  the same elliptical
annulus as NEdge$_{+1}$ but is constrained to lie 
between $213^\circ$ and $327.5^\circ$ (to the south). All angles are
measured counter-clockwise from west. Coordinates are
J2000. $^a$ Radii are semi-minor inner (outer) radii for elliptical 
annular sectors. 
} 
}
\end{deluxetable}

\begin{deluxetable}{ccccccc}
\tablewidth{0pc}
\tablecaption{NGC\,6868 Spectral Models Across the Edge\label{tab:n6868edgespec}}
\tablehead{\colhead{region} &\colhead{counts} & \colhead{$kT$} & \colhead{$A$} &
\colhead{$\Gamma$}  & \colhead{$\chi^2/{\rm dof}$}& \colhead($\Lambda^a$) \\
   &(keV) & ($\Zs$) &  & & &  
}
\startdata
NEdge$_{-1}$ &$313$ &$0.65^{+0.6}_{-0.7}$ & $0.3$ & \ldots &$11.6/19$ & $6.2$ \\
NEdge$_{-1}$ &$313$ &$0.65 \pm 0.07$ &$0.5$  &\ldots  & $11.2/19$ & $9.2$\\
NEdge$_{-1}$ &$313$ &$0.64 \pm 0.08$&$0.5$  &$1.6$  & $10.5/18$ & $9.2$ \\
NEdge$_{-1}$ &$313$ &$0.66 \pm 0.06$&$1.0$ & \ldots & $11.1/19$   & $16.8$ \\
NEdge$_{+0}$ &$286$ &$0.71^{+0.08}_{-0.11}$ & $0.2$ & \ldots & $30.2/31$ & $4.7$ \\
NEdge$_{+0}$ &$286$ &$0.72^{+0.08}_{-0.11}$&$0.3$&\ldots& $30.6/31$ &$6.1$ \\
NEdge$_{+1}$ &$464$ & $0.73^{+0.07}_{-0.09}$&$0.3$&\ldots &$65.7/55$ & $6.1$
\\
NEdge$_{+1}$ &$464$ &$0.69^{+0.08}_{-0.10}$&$0.11^{+0.13}_{-0.10}$
&\ldots & $61.7/54$ & $3.3$ \\
S$_{+1}$ &$691$ &$0.63^{+0.05}_{-0.06}$&$0.29^{+0.24}_{-0.18}$
&\ldots &$72.1/68$ & $6.1$ \\
\enddata
\tablecomments{\footnotesize{Spectral models are an absorbed APEC
    model or absorbed APEC + power law model. Hydrogen absorption is
    fixed at the Galactic value (N$_{\rm H} = 3.9 \times 10^{20}$\cms, 
    Kalberla \etal 2005). Parameters without 90\% CL uncertainties 
    are fixed. $^a \Lambda$ is the X-ray emissivity in the 
   $0.5-2$\,keV energy band in units of 
   $10^{-15}\,{\rm photon\,cm}^3\,{\rm s}^{-1}$. }
} 
\end{deluxetable}

\begin{deluxetable}{ccccccc}
\tablewidth{0pc}
\tablecaption{NGC\,6868 Edge Analysis\label{tab:n6868mach}}
\tablehead{\colhead{case$^a$} & \colhead{$T_1/T_2$} & 
\colhead{$(\Lambda_1/\Lambda_2)^{1/2}$} &\colhead{$n_1/n_2$} & \colhead{$p_1/p_2$} &
\colhead{Mach} &\colhead{velocity} \\
& & & & & & (\kms) 
} 
\startdata
$1$&$0.89$ &$0.99$ &$1.67-2.56$ &$1.49-2.28$ &$0.72-1.08$ &$318-475$\\
$2$&$0.88$ &$0.81$ &$1.37-2.10$ &$1.21-1.85$ &$0.47-0.9$ &$207-397$ \\
$3$&$0.90$ &$0.60$ &$1.01-1.55$ &$0.90-1.39$ &$0 - 0.65$ &$0 - 287$ \\
\enddata
\tablecomments{\footnotesize{Case ($1$,$2$,$3$) correspond to spectral models 
given in Table \protect\ref{tab:n6868edgespec} with  abundances 
inside the edge of ($0.3$,$0.5$,$1.0$), respectively, and temperature, 
abundance and sound speed outside the edge of $0.72$\,keV, $0.3\,\Zs$
and $440$\kms. 
}}
\end{deluxetable}

\begin{deluxetable}{ccccc}
\tablewidth{0pc}
\tablecaption{NGC\,6868 Spectral Regions for Temperature Map Features\label{tab:n6868tmapspecreg}}
\tablehead{\colhead{label} & \colhead{shape} & \colhead{center} &
  \colhead{radius$^a$}&\colhead{angle} \\
& & (RA, Dec) & (arcsec) & (deg) 
}
\startdata
T$^b$&ellipse & $20:10:13.3,-48:27:16.1$ &$196,105$&$333$ \\
     & ellipse &  $20:09:40.0,-48:27:02.4$  & $149.6,92.8$& $21.5$ \\
F & ellipse  &$20:09:45.4,-48:24:27.0$& $150.,62.7$& $29.2$ \\ 
C & circle  &$20:10:04.9,-48:22:54.4$  &$72.5$& \ldots  \\ 
N & circle  &$20:09:53.7,-48:22:46.6$ & $6.2$ &\ldots  \\ 
\enddata
\tablecomments{\footnotesize{All angles are
measured counter-clockwise from west. Coordinates are
J2000.   $^a$ radii  are semi-(major,minor) axes for ellipses and radius
for circles  $^b$ T is the sum of the two elliptical regions given.  
} 
}
\end{deluxetable}

\begin{deluxetable}{cccccc}
\tablewidth{0pc}
\tablecaption{NGC\,6868 Spectral Models for Temperature Map Features\label{tab:n6868tmapspec}}
\tablehead{\colhead{region} &\colhead{counts} & \colhead{$kT$} & \colhead{$A$} &
\colhead{$\Gamma$}  & \colhead{$\chi^2/{\rm dof}$} \\
   &(keV) & ($\Zs$) &  & & 
}
\startdata
T &$785$ &$0.62 \pm 0.04$  &$0.4^{+1.07}_{-0.20}$ &\ldots & $79.0/79$ \\
F &$607$ &$0.71 \pm 0.05$ &$0.3$ &\ldots & $37.3/37$ \\
F &$607$ &$0.70 \pm 0.05$ &$0.2$ &\ldots & $35.2/37$ \\
C &$504$ &$0.59^{+0.04}_{-0.05}$ &$0.5$ &\ldots &$38.2/28$ \\
N &$790$ &$0.63^{+0.07}_{-0.06}$ &$1.0$ &$1.6$ &$37.7/31$ \\
N &$790$ &$0.63^{+0.07}_{-0.06}$ &$1.0$ &$1.4$ &$32.1/31$ \\
N &$790$ &$0.63^{+0.07}_{-0.06}$ &$1.0$ &$2.0$ &$51.9/31$ \\
\enddata
\tablecomments{\footnotesize{Spectral models are an absorbed APEC
    model or absorbed APEC + power law. Hydrogen absorption is
    fixed at the Galactic value N$_{\rm H} = 3.9 \times 10^{20}$\cms, 
    (Kalberla \etal 2005). Parameters without 90\% CL uncertainties 
    are fixed.}
} 
\end{deluxetable}

\begin{deluxetable}{ccccc}
\tablewidth{0pc}
\tablecaption{NGC\,6861 Spectral Regions\label{tab:n6861specreg}}
\tablehead{\colhead{label} & \colhead{shape} & \colhead{center} &
  \colhead{radius$^a$}&\colhead{angle} \\
& & (RA, Dec) & (arcsec) & (deg) 
}
\startdata
E  &rectangle& $20:07:51.8,-48:22:01.0$&$284.7,392.3$ & $0$  \\
G  & ellipse  &$20:07:19.5,-48:22:12.9$ & $31.6,19.4$ &$33.8$   \\ 
FA$^b$ & annular sector  & $20:07:19.5,-48:22:12.9$&$39.9,106.8$ & \ldots  \\ 
WT & rectangle & $20:07:01.6,-48:22:13.7$&$281.4,269.6$ & $0$ \\
NW & rectangle  &$20:07:14.3,-48:17:46.5$ &$372.8,147.2$ &$47.3$ \\
\enddata
\tablecomments{\footnotesize{ Angles are
measured counter-clockwise from west. Coordinates are
J2000.   $^a$ radii  are (length, width) for rectangles, 
semi-(major,minor) axes for the ellipse, and (inner,outer) radius
for the annular sector. 
$^b$ annular region is constrained to lie between 
$80^\circ$ and $282^\circ$ measured counterclockwise from
west.  
} 
}
\end{deluxetable}

\begin{deluxetable}{cccccc}
\tablewidth{0pc}
\tablecaption{NGC\,6861 Spectral Models\label{tab:n6861specfits}}
\tablehead{\colhead{region} & \colhead{source} &\colhead{$kT$} & \colhead{$A$} &
\colhead{$\Gamma$} &  \colhead{$\chi^2/{\rm dof}$} \\
 & counts  &(keV) & ($\Zs$) &  & 
}
\startdata
E1  & $404$ & $0.61^{+0.24}_{-0.07}$ &$0.027^{+0.04}_{-0.02}$ &\ldots &$28.5/22$ \\
E2  & $404$ &$1.26^{+0.44}_{-0.26}$ &$0.2$ &  & \\
   &       & $0.52^{+0.17}_{-0.18}$& $0.2$ &\ldots &$26.3/21$ \\
G &$436$ &$0.66^{+0.11}_{-0.07}$ &$0.5$ &$1.86^{+0.57}_{-0.55}$ & $8.6/16$ \\
FA &$214$ &$1.3^{+0.3}_{-0.2}$ &$0.5^{+1.8}_{-0.28}$ &\ldots & $10.6/9$\\
WT & $621$  &$1.18^{+0.10}_{-0.12}$ &$0.25^{+0.21}_{-0.13}$ &\ldots & $35.7/42$ \\
NW  & $385$ &$0.90 \pm 0.09$ & $0.22^{+0.26}_{-0.11}$ & \ldots & $39/28$ \\
\enddata
\tablecomments{\footnotesize{Spectral models are an absorbed APEC
    thermal plasma model, two component absorbed APEC or absorbed 
    APEC + power law. Hydrogen absorption is
    fixed at the Galactic value N$_{\rm H} = 3.9 \times 10^{20}$\cms, 
    (Kalberla \etal 2005). Parameters without 90\% CL uncertainties 
    are fixed.}
} 
\end{deluxetable}

\vfill
\eject
\end{document}